\documentstyle[epsfig,rotate]{aa}
 
\begin{document}
 
\thesaurus{02.08.1; 03.13.4; 02.19.1; 02.07.1;}
\title{Time-Independent Gravitational Fields in the BGK Scheme
for Hydrodynamics}

\author{Adrianne Slyz\inst{1,2} \and Kevin H. Prendergast\inst{1}}
\offprints{slyz@mpia-hd.mpg.de}
\institute{Columbia University,
           10027 New York, USA
\and       Max-Planck Institut f\"{u}r Astronomie,
           K\"{o}nigstuhl 17, D-69117 Heidelberg, Germany
}
\date{Received ???; accepted ???}
\authorrunning{A. Slyz \& K.H. Prendergast}
\titlerunning{BGK with Gravity}
\maketitle
 
\begin{abstract}
We incorporate a time-independent gravitational
field into the BGK scheme for numerical hydrodynamics.  In the 
BGK scheme the gas
evolves via an approximation to the collisional Boltzmann equation, namely
the Bhatnagar-Gross-Krook (BGK) equation.  Time-dependent
hydrodynamical fluxes are computed from local solutions of the 
BGK equation.  By accounting
for particle collisions, the fundamental mechanism for generating
dissipation in gas flow, a scheme based on the BGK equation gives
solutions to the Navier-Stokes equations: the fluxes carry both
advective and dissipative terms.  We perform numerical experiments
in both 1D Cartesian geometries and axisymmetric cylindrical coordinates.
\keywords{hydrodynamics---methods:numerical---shock waves---gravitation}
\end{abstract}
\section{Introduction}
\label{sec-intro}

The BGK code distinguishes itself from
other hydrocodes in that it has recourse 
to the physics which generates
dissipation, namely the physics of particle collisions.
Developed by Prendergast \& Xu (1993), the hydrodynamical scheme 
invokes a microscopic description of gas flow and it is therefore
based on considerations
of gas kinetic theory.
Recall that kinetic-based hydrocodes rely on the
fact
that the state of a gas can be described by giving the distribution
function of particle velocities $f({\bf x},{\bf u},t)$ at a point in the phase space
of a single particle.   The codes take advantage of the fact that
the quantities of hydrodynamical interest (namely the mass, momentum and 
energy densities in the gas) are low-order velocity moments of $f$.   
$f$ obeys the Boltzmann-equation, which we write as 
$\frac{Df}{Dt} = {\frac{\delta f}{\delta t}}^{coll}$, where $\frac{D}{Dt}$ 
is a time rate of
change along the  trajectory of a single  particle 
moving freely in phase space (under the action of smoothly varying
forces, if these are present) and $\frac{\delta f}{\delta
t}^{coll}$ is the rate of change of $f$ due to collisions. 
The classical two-body  collision integral is non-linear, and
non-local in velocities (Cercignani 1988). 
BGK (from the names of Bhatnagar-Gross-Krook) replaces this integral with the term $\frac{(g-f)}{\tau}$, 
where $g$ is the
Maxwell-Boltzmann distribution function having the same mass, momentum and
energy densities as $f$, and $\tau$ is a relaxation time, which can (ideally)
be as small as the mean time between collisions of a particle in the gas
(Bhatnagar, Gross \& Krook 1954).
It is important to notice that $f$ and $g$ have the same mass,
momentum and
energy densities, but they do not give rise to the same fluxes of these
quantities.  Fluxes are determined by different moments of the distribution 
functions.  
BGK differs from all other kinetic-based codes
 in several respects: the collisions are active throughout
the duration of a time step, and are not imposed as a distinct
``equilibration'' process at the end of a timestep. Also, we do not guess
the form of the distribution function but solve the BGK equation to find
it.         
The BGK equation is linear in $f$, which makes it easy to solve
if $g$ is known.  However $g$ is not known; the ``compatability''
 conditions that
$f$ and $g$ have the same mass, momentum and energy densities would determine
$g$ if $f$ were known. Therefore the BGK formulation is also nonlinear and
non-local in velocity space, as is the Boltzmann equation. It will be
shown that this situation leads to a set of non-linear integral equations
for the parameters of g.    
It might seem that we have made no progress by replacing the
Boltzmann equation by the BGK plus compatability conditions; but this is not
true, because it will be shown that we need solve for the parameters of g
only in the neighborhood of a boundary between computational cells, and
for a short time (given by the usual CFL (Courant-Friedrichs-Lewy) condition).  Knowledge of the
parameters of $g$ is equivalent to knowledge of the mass, energy and
momentum densities.            

The connection between the BGK equation's microscopic description of gas flow
and a macroscopic description 
has been shown by
Cercignani (1988) 
for the case of a perfect monatomic gas and by Xu (1993) for polyatomic
gases. Velocity moments of the BGK equation give the Euler equations 
for negligible particle collision
time, $\tau$, the Navier-Stokes equation for small yet non-zero $\tau$
and a description of rarefied gas dynamics for large $\tau$.
In the Navier-Stokes regime, these derivations also furnish 
expressions for the shear and
bulk viscosity coefficients and the heat conductivity coefficient.  

In principle, if the local value of the collision time can be measured
as a function of time for a gas, then the BGK scheme may be used to
evolve the gas with true physical dissipation parameters.
In practice, the resolution of the BGK scheme is limited by the
grid resolution.  If the grid is not fine enough
to resolve a discontinuity, then artificial dissipation must be added
to broaden the discontinuity so that it is
at least one grid cell thick.  Because viscosity and
heat conductivity are proportional to $\tau$, the BGK scheme broadens 
shocks by enlarging $\tau$ at the location of the discontinuities.  
The expression for the collision time in the BGK scheme therefore
contains two terms. One term is the real physical mean collision 
time and it is chosen according to the desired Reynolds number of 
the problem.  The second
term is chosen in such a way that shocks in the flow span
at least one grid cell.  The latter term tunes the amount of artificial
dissipation in the scheme.  The notable difference between how the
BGK scheme inputs artificial dissipation and how other schemes input
artificial dissipation is that the BGK scheme puts it in exactly as
if it were real dissipation corresponding to the numerically necessary
value for $\tau$.

We emphasize that modelling the real dissipation in an astrophysical object
such as that resulting from ``turbulent'' viscosity, is still outside 
the reach of any existing hydrocode.  For one thing, values for
viscosities in astrophysical objects are highly speculative.
Secondly, to solve Navier-Stokes problems with a particular value
of the 
viscosity for an astrophysical object requires grid
resolutions which are beyond what is currently achievable.
For laboratory scale phenomena and for small enough Reynolds number,
the BGK code is successful at
modelling the dissipation produced by particle collisions in 
a fluid. Tests of the Kolmogorov and the laminar boundary layer
problem show real dissipative effects (Xu \& Prendergast 1994).

Thus far the BGK code has been extensively tested without the inclusion of
gravity.  Several papers document the tests which verify
its accuracy and robustness as both an Euler and Navier-Stokes solver
in 1D and 2D Cartesian geometries and for two-dimensional adaptive
unstructured grids.  The list of one-dimensional Euler tests which 
have been performed with the BGK scheme includes: the Roe, Sod, Lax-Harten,
Woodward-Colella, and Sj{\"{o}}green tests for subsonic and
supersonic expansion (Prendergast \& Xu 1993; Xu 1993; Xu, Martinelli \&
Jameson 1995). The results from these
one-dimensional test cases are that the BGK scheme produces shock fronts which
are typically one to two cells wide, and contact discontinuities which
are slightly broader.  This is competitive with
high resolution codes which do not employ regridding in the
neighborhood of a shock front.  The BGK scheme also exhibits negligible 
under and over shooting even at strong shock fronts.  The other
notable features in the tests with rarefaction waves are the sharp
corners at the junctions between the rarefaction waves and the
undisturbed, uniform regions.  

Many codes have trouble treating a low density region with a flow
gradient. The BGK scheme  successfully handles low density regions  
because it satisfies both an entropy condition and a positivity condition
(Xu {\rm et al.} 1996).  
In the BGK scheme collisions relax the gas
toward local thermodynamic equilibrium states and this relaxation
process is accompanied by an increase in entropy.  
Godunov-type schemes on the other hand typically demand an 
entropy fix when they encounter strong rarefaction waves ({\rm cf.} 
the Sj{\"{o}}green test) otherwise they produce unphysical rarefaction shocks.
Usually the fix is an addition of artificial viscosity.
Because the BGK scheme naturally satisfies the entropy condition, 
it simply cannot generate these unphysical phenomena.  
In satisfying the positivity condition, the BGK scheme 
avoids producing states with negative
density or internal energy.  The Roe test is
conducive  to the creation of these non-physical states but the BGK scheme
does not encounter them (Prendergast \& Xu 1993).  The same cannot be 
said of the performance
of  conservative finite difference schemes and codes
employing Riemann solvers ({\rm e.g.}  Roe's approximate Riemann
solver).

The list of two-dimensional Euler test cases which have been
performed with the BGK scheme includes: (a) uniform Mach 3 flow 
in a tunnel with a forward facing step (the Emery test).  The BGK code
is not modified in any way near the step to treat the flow past it and
its corner, which is a singular point.  With the BGK scheme expansion
shocks never emerge from the corner (Prendergast \& Xu 1993; Xu 1993;
Xu {\rm et al.} 1995), (b) double Mach
reflection in supersonic flow over a wedge, (c) the diffraction of a
strong shock (${\rm Mach} = 5.09$) around a corner. Test cases (b) and (c)
are further examples of
the BGK scheme succeeding at simulating flow without having to
summon  any detection algorithms or
entropy fixes.  According to Xu {\rm et al.} 1995, the original
Godunov scheme, the Roe scheme without the entropy fix, and the Osher
scheme could produce a rarefaction shock at the corner in test case (c); (d) flow around an impulsively started
cylinder.  When applied to this problem, many schemes either fail or
have severe problems in maintaining positive pressure and density
in the near vacuum low pressure and low density region created behind
the cylinder.  The BGK scheme seems to be able to preserve positivity
without {\em ad hoc} fixes, and to reach a steady state solution for
the problem (Xu {\rm et al.} 1996). 

To test its performance as a Navier-Stokes solver and to see if
it has real viscosity effects, the BGK scheme has
been applied to the laminar boundary layer problem and the Kolmogorov 
problem (Xu 1993).
The laminar boundary layer problem models the flow of gas above 
a flat plate.  Even on coarse grids ({\rm e.g.} (32 X 16), (16 X 8)),
the BGK scheme impressively recovers the Blasius profile.
In the Kolmogorov problem, 
a one-dimensional sinusoidal velocity
field is imposed in a uniform density and isothermal fluid.  The BGK
code fulfills the Navier-Stokes prediction which is that the shape of
the fluid's velocity profile remains unchanged while the amplitude of
the velocity decreases in such a way that the fluid's kinetic energy
decays exponentially.  The agreement between the
theoretical viscosity coefficient and the numerical viscosity
coefficient (deduced from the measured decay rate) is excellent.  

To improve the resolution of physical discontinuities occurring in
complicated flows, a version of the BGK scheme on a two-dimensional
adaptive unstructured grid has now been developed (Kim \& Jameson
1998).  

Most astronomical applications of a hydrocode require a consideration
of gravity.  To this end we have been developing the BGK scheme.  
Recently the BGK scheme has been used for a cosmological simulation
(Xu 1997) but this is prior to results showing the long-term stability
of the BGK scheme with gravity and the BGK scheme's convergence to the 
equilibrium state with gravity.  In this paper we give such results.
In addition to the incorporation of gravity, we have modified the geometry of
the Eulerian grid to axisymmetric cylindrical coordinates.  
Changing the geometry of the grid has effects similar to the
effects of adding gravity to the BGK scheme.  It gives rise to
source terms in the hydrodynamic equations.  For the purpose 
of simplicity, in this paper we 
describe modifications to 
the BGK method when a time-independent gravitational potential 
is incorporated into a BGK scheme
in Cartesian coordinates (section~\ref{sec-fsg}).  
We present results however for both
gas flow on a one-dimensional Cartesian grid and on an axisymmetric
cylindrical grid (section~\ref{sec-results}).

\section{A BGK Flow Solver with Gravity}
\label{sec-fsg}
\subsection{Hydrodynamic Equations from the BGK Equation with Gravity}
\label{sec-hydeq}
In Cartesian coordinates, we specify the position and
velocity of a particle with coordinates $({\bf x},{\bf u})$.  
To be clear, ${\bf x}$ here is a 3-vector $(x,y,z)$ giving the
position of a particle in configuration space and ${\bf u}$ 
is a 3-vector $(u,v,w)$ giving the velocity of a particle in this space.
In this coordinate
system and in the presence of a smooth gravitational field with
gravitational potential, $\Phi({\bf x})$,  the
BGK equation is:
\begin{eqnarray}
\frac{\partial f}{\partial t}  +  {\bf u}\cdot \frac{\partial f}{\partial {\bf x}}
-  \frac{\partial{\Phi}}{\partial {\bf x}} \cdot
\frac{\partial f}{\partial {\bf u}} = \frac{g - f}{\tau}.
\label{eq:cylBGK}
\end{eqnarray}
(For reference, this is equivalent to equation 2.15 in Shu (1992) where
${\frac{\delta f}{\delta t}}^{coll}$ is the BGK collision term 
$\frac{g - f}{\tau}$.)
Recall that in the BGK collision term $g$ is the equilibrium
distribution  towards which the true velocity distribution
function $f$ relaxes on a collision timescale, 
$\tau$.  We define the units of $f$ and $g$ to be the mass
in a volume element of phase space.
The particle collision time $\tau$ depends on macroscopic quantities
such as temperature and density, and it is therefore a function
of position ${\bf x}$ and time $t$.
Both $f$ and $g$ are functions of space ${\bf x}$, time $t$, and particle
velocities ${\bf u}$ and $\xi$ where  
$\xi = (\xi_{1},\xi_{2}, \ldots \xi_{K})$ is a $K$-dimensional vector of
velocities associated with the internal degrees of freedom of a particle.
With this parameterization of the internal velocities,
the energy associated with internal degrees of
freedom is written as $\frac{1}{2} m \xi^{2}$ where 
${\xi}^{2}={\xi_{1}}^{2}+{\xi_{2}}^{2}+ \ldots + {\xi_{K}}^{2}$.
Included in these internal degrees of freedom is
the energy not explicitly accounted for in versions of the code with
physical spatial dimension, D, less than three.  For example a
particle in a diatomic gas $(\gamma = 7/5)$ moving in 3 spatial
dimensions has 5 degrees of freedom.  (The connection between the
number of degrees of freedom, $n$, and the ratio of specific heats,
$\gamma$, is $\gamma = (n + 2)/n$.)  Because a one dimensional hydrocode
$(D = 1)$ treats only one of these degrees of freedom explicitly, 
the behavior of the diatomic gas is reproduced by treating the four
remaining degrees of freedom as internal energies.  We represent 
those energies
in the K-dimensional vector $\xi$ where $K = n - D$.  In terms of
$\gamma$, $K = \frac{2}{(\gamma - 1)} - D$.

Equations governing the time evolution of the mass density, 
momentum density and kinetic plus internal energy density 
are found by multiplying 
(\ref{eq:cylBGK}) in turn by 1, 
${\bf u}$, and $\frac{1}{2}({\bf u}^{2}+\xi^{2})$
and integrating over a volume in velocity space, $d\Xi =
d{\bf u}\,\xi^{(K-1)} \, d\xi$.  
The volume element in
velocity space has the above form in the $\xi$ variable 
because we need only the total internal energy of
a particle and not the energy associated with any particular internal
degree of freedom.  Hence an integration has  been performed
over angles in $\xi$.  
An essential point about the derivation of 
the hydrodynamic equations from velocity moments of the 
BGK equation is that moments of the BGK
collision term, $\frac{( \mbox{\em g}-\mbox{\em f})}{\tau}$, vanish.
This is enforced by the fundamental 
hypothesis of the BGK model: in the relaxation process 
collisions tend to reshuffle the particles defining the true 
distribution function, 
$f({{\bf x}}, {{\bf u}},t)$ into a local equilibrium distribution, 
$g({{\bf x}}, {{\bf u}},t)$ having the
same mass, momentum and energy densities as $f({{\bf x}}, {\bf{u}},
t)$.  We call this requirement for the
equivalence of $f$ and $g$'s moments, the ``compatability condition''.

For detailed results of taking velocity moments of (\ref{eq:cylBGK}), 
we refer once again to Shu 1992 (p.20-23).
We give the moment equations here in compact form:
\begin{equation}
\frac{\partial}{\partial t} \langle \zeta \rangle + 
\frac{\partial}{\partial x_{k}} \langle u_{k} \zeta \rangle + 
\frac{\partial \Phi}{\partial x_{k}} \langle {\frac{\partial \zeta}
{\partial u_{k}}}\rangle = 0
\label{eq:bracketmom}
\end{equation}
where we have defined the following bracket notation:
\begin{displaymath}
\langle \ldots \rangle=\int (\ldots) \,f \, d\Xi
\end{displaymath}
and $\zeta = 1, u_i, \frac{1}{2}({\bf u}^{2}+\xi^{2})$.
In eq.~\ref{eq:bracketmom} we see that mass, momentum and energy
densities are updated by terms which are the divergence of a quantity
(i.e. this quantity is commonly referred to as a flux) and 
when gravity (or a curved coordinate system) is involved, 
also by terms which cannot generally
be written as the divergence of a quantity and hence cannot
update the contents of a cell by a flux through the cell's surface. 
The latter terms are called source terms.  
Without gravity and in Cartesian coordinates, 
a BGK scheme for hydrodynamics evolves all the hydrodynamic quantities
exclusively by flux terms.
We emphasize that there is no dissipation source term in equation~\ref{eq:bracketmom}. Dissipative terms are carried by the fluxes.  The BGK scheme 
is unique in this
respect.  In Cartesian coordinates it allows a fluid to evolve concurrently through advective and
dissipative processes without decoupling them into two separate
operations.  

With gravity, source term computation is unavoidable.
The gravitational source term in the momentum equation
cannot be manipulated into the form of the divergence of some
quantity, so it persists as a source term.
The gravitational source term in 
the energy equation on the other hand may be reformulated to 
give an energy equation
without a source term if
the gravitational field is independent of time.
In this paper  we consider this case.
Without a gravitational source term, the energy equation for the
case of a time-independent gravitational field is a
conservative equation for the total kinetic, internal and
gravitational energies.  In one dimension it is:
\begin{displaymath}
\frac{\partial}{\partial t} \left({\cal E}_{\rm kin} + {\cal E}_{\rm int} 
+ \rho \Phi \right) +
+ \frac{\partial}{\partial x} \left( 
{ F}^{{\cal E}_{\rm kin}+{\cal E}_{\rm int}} +
\Phi {{F}}^{\rho}
\right) =0 .
\label{eq:consener}
\end{displaymath}
We define $F^{{\cal E}_{\rm kin}+{\cal E}_{\rm int}}$ here as the flux of the kinetic plus internal
energy, and $F^{\rho}$ as the mass flux.  In $F^{{\cal E}_{\rm kin}+{\cal E}_{\rm int}}$ and $F^{\rho}$
we are only considering the $x$-component of the flux, i.e. the flux
through a wall perpendicular to the $x$ direction.

In the end, the BGK scheme follows the time evolution of the
integrated  values of mass, momentum and energy densities within cells. 
For example, for one dimensional flow in Cartesian
coordinates the BGK scheme tracks the contents of a volume element 
with boundaries $x_{1}$ and $x_{2}$. 
If the values of mass, momentum
and energy densities (denoted below by ${{q}}$) are given within 
this cell at time $t_{1}$ then
at time $t_{2}$ the new values of the mass, momentum and energies are
updated by fluxes (${\cal{F}}$) (section~\ref{sec-hydfluxesg}) through 
the cell's 
boundaries and in some cases by source terms (${{S}}$)
(section~\ref{sec-src}).
\begin{eqnarray}
\int^{x_{2}}_{x_{1}} \, {q}({x},t_{2}) \, d {x} = 
\int^{x_{2}}_{x_{1}} \, {q}({x},t_{1}) \, d {x}
 & + & \nonumber \\ {\int^{t_{{\rm 2}}}_{t_{{\rm 1}}}} \, {\cal{F}}{({x_{{\rm 1}}},t) \cdot \hat{\bf{x}}\, d t}
- {\int^{t_{{\rm 2}}}_{t_{{\rm 1}}}} \, {\cal{F}}{({x_{{\rm 2}}},t)\cdot \hat{\bf{x}} \, d t} & + & \nonumber \\ {\int^{t_{{\rm 2}}}_{t_{{\rm 1}}}} \int^{x_{2}}_{x_{1}} \, {S}({x},t) \, d {x} \, d t .
\label{eq:intconserv}
\end{eqnarray}

\subsection{Hydrodynamical Fluxes with Gravity}
\label{sec-hydfluxesg}

Fluxes in the BGK scheme arise from velocity moments of a particle
distribution function $f$ which is
a solution to the ordinary differential equation 
$\frac{D f}{D
t} + \frac{f}{\tau} =  \frac{g}{\tau}$ with initial conditions $f =
f_{0}$ at $t = t_{0} = 0$.  This equation holds along each trajectory.
The solution is composed of two terms: an integral over the past
history of $g$ and a term representing relaxation from an initial
state, $f_{0}$.  
\begin{eqnarray}
f({\bf x}_{f},{\bf u},t)=\frac{1}{\tau} \int^t_{0}
g({{\bf x}}^{'},{{\bf u}}^{'},
t^{'}) e^{-(t-t^{'})/\tau} \, dt^{'} + \nonumber \\
e^{-t/\tau} \, f_{0}({\bf x}_{f} - {\bf u} t,{\bf u},t_{0})
\label{eq:formalBGKsoltn}
\end{eqnarray}
Here ${\bf x}^{'}$ and ${\bf u}^{'}$ in the arguments of $g$ are 
solutions of a gas particle's equations
of motion in Cartesian coordinates with gravity:
\begin{eqnarray*}
\frac{d {\bf x}^{'}}{dt^{'}}  =  {\bf u}^{'};  &     & \frac{d {\bf u}^{'}}{dt^{'}}  = 
 - \frac{d \Phi}{d {\bf x}^{'}} 
 \nonumber \\
\end{eqnarray*} 
with the final conditions ${\bf x}^{'} = {\bf x}_{f}$ and
${\bf u}^{'} = {\bf u}$ at $t^{'}=t$.
To first order in time the solutions of these equations are:
\begin{eqnarray}
{\bf x} ^{'} = {\bf x}_{f} - {\bf u} (t-t^{'});  &     & {\bf u}^{'} = {\bf u} -
\left(-{\frac{d \Phi}{d {\bf x}}}
\right)_{t^{'}=t}
(t-t^{'}).
 \nonumber \\
\label{eq:soltnseoms}
\end{eqnarray} 
These equations define a single trajectory for the particles which 
arrive at ${\bf x}_{f}$ at time $t$ with velocity ${\bf u}$.  The
trajectory would generally be curved if we carried terms of higher
order than those shown.
Even when we are formulating the BGK scheme for a 1-dimensional
Cartesian grid there is no escape from considering $\bf{u}$ a vector
in three dimensions.  Both $f$ and $g$ always depend on all of the
``molecular'' velocity components because individual particles
can have velocities of any speed and in any direction.
A consequence of this is that when fluxes are eventually 
computed at ${\bf x}_{f}$, $f({\bf x}_{f},{\bf u},t)$  
is integrated over the full and continuous
range of velocities, ${\bf u}$, from $-\infty$ to $+\infty$.  
Therefore fluxes computed from the
true distribution function $f({\bf x}_{f}, {\bf u}, t)$ arise
by a weighted integration of the equilibrium distribution function
$g({{\bf x}^{'}}, {{\bf u}^{'}}, t^{'})$  not just over one trajectory but 
over all possible trajectories (in 6-dimensional phase space)
which arrive at (${\bf x}_{f},
t$). (We leave it to section~\ref{sec-implication} for a discussion
of the implications of this.)  For a 1-dimensional problem in Cartesian 
coordinates all we
require is that the trajectory wind up on the wall located at $(x_{f},y_{f},z_{f})$ where $y_{f}$ and $z_{f}$ are fixed for all $x$ because the only non-trivial
dimension is the $x$-dimension.

Now as already stated in the introduction,
$g$ is not known {\em a priori} along all the  trajectories passing through
${\bf x}_{f}$ at time t.
Physically $g$ must describe an equilibrium state so 
it is sensible to assume a Maxwellian form for
$g$:
\begin{equation}
g({\bf x}^{\prime}, {\bf u}^{\prime}, t^{\prime})=\rho
\left(\frac{\lambda}{\pi}\right)^{(K+3)/2} 
e^{- \lambda \left(({\bf u}^{\prime}-{\bf U})^{2}+\xi^2 \right)}.
\label{eq:maxwellian}
\end{equation}
Here ${\bf U} = (U,V,W)$ are the mean macroscopic velocities in
the $x$, $y$ and $z$ directions respectively.  To explain why ${\bf U}$
has this physical interpretation we go to the one-dimensional problem.
In one-dimension, $V$ and $W$ are zero so $g$ is effectively left with
3 parameters: $\rho$, $U$ and $\lambda$.  
For a perfect gas with $K = 2/(\gamma - 1) - D$ the ``compatability''
condition gives an explicit relationship in closed form between
the mass, momentum and energy densities and $g$'s parameters:
\begin{equation}
\left(\begin{array}{clcr}
\rho \\ P_{x} \\
\cal{E}_{\rm kin} + \cal{E}_{\rm int}
\end{array}\right) = 
\left(\begin{array}{clcr}
\rho \\ \rho U  \\ 
\frac{\rho}{2}(U^{2}+\frac{(K+3)}{2 \lambda})
\end{array}\right).
\label{eq:parcorr}
\end{equation}
(Note that when $g$ is Maxwellian its velocity moments follow
simple recurrence relations (see appendix in Xu {\rm et al.} (1996)).)
Eq.~\ref{eq:parcorr} assigns physical meaning to 
$g$'s parameters:
$\rho$ is the mass
density, $U$ is the mean macroscopic velocity in the
${x}$ direction, and 
$\lambda$ is related to the mass density, $\rho$, and the internal 
energy density, $\cal{E}_{\rm
int}$,  in the following way:
\begin{displaymath}
\lambda = \frac{(K+3)}{4} \frac{\rho}{\cal{E}_{\rm int}}
\end{displaymath}
where $K$ was defined earlier as the dimension of the vector $\xi$.
$\lambda$ is therefore inversely proportional to the temperature of
the gas.
Of course mass densities, mean macroscopic velocities and internal
energy densities change with space and time in the fluid and 
their evolution is assumed to be governed by the evolving
distribution function $f$.  Hence there is a mutual connection
between $f$ and $g$.
By substituting the formal solution for $f$ 
(eq. \ref{eq:formalBGKsoltn})
which itself depends on $g$ into the ``compatability condition'' on the
moments of $g$, one arrives at a
set of coupled
non-linear integral equations for the space and time dependence of
$g$'s parameters: 
\begin{eqnarray}
\int {\psi_{\alpha}} e^{-\chi_{\alpha} \psi_{\alpha}({\bf r},t)} d \Xi =
\nonumber \\
\frac{1}{\tau} \int {\psi_{\alpha}} \int^{t}_{0} e^{-\chi_{\beta}
\psi_{\beta}({ {\bf r}^{\prime}},t^{\prime}) -
\frac{(t-t^{\prime})}{\tau}}
d t^{\prime} d \Xi  
\nonumber \\
+ e^{-t/\tau} \int {\psi_{\alpha}} f_{0}({\bf 
x}_{f}-{\bf u}t,t_{0}) d \Xi .
\label{eq:coupinteqs}
\end{eqnarray}
Here we have  adopted the following abbreviated notation for 
$g$ and its parameters.
Let
\begin{displaymath}
g = e^{- \chi_{\alpha} \psi_{\alpha}}  
\end{displaymath}
where $\alpha = 1, \ldots, 5$ and
\begin{displaymath}
\psi_{\alpha} = ( 1, {\bf u}^{\prime}, \frac{1}{2}({{\bf u}^{\prime}}^{2}+\xi^{2})).
\end{displaymath}
We define the transpose of ${\psi_{\alpha}}$ to be ${\psi_{\alpha}}^{\rm T}$
and $\chi_{\alpha}$ to be a vector of $g$'s parameters:
\begin{displaymath}
\chi_{\alpha} = ( -ln[\rho (\lambda/\pi)^{(K+3)/2}],-2\lambda U,-2\lambda V, 
-2\lambda W, 2\lambda).
\end{displaymath}
Notice that ${\psi_{\alpha}}$ has no space dependence whereas $\chi_{\alpha}$
has no velocity dependence.
In equation (\ref{eq:coupinteqs}) we assume that 
the collision time, $\tau$, which is
a function of macroscopic densities is locally constant and hence can
be removed outside the integral which is performed over the velocity moments.  
The computation of $\tau$ is discussed in section~\ref{sec-collisiontime}.
Because $f_{0}$ in the formal
solution for $f$ (eq.\ref{eq:formalBGKsoltn}) may be straighforwardly 
constructed from the 
initial conditions, and because
the integral over $g({\bf x}^{\prime}, {\bf u}^{\prime},
t^{\prime})$ in the formal solution for $f$ is readily evaluated 
once $g({\bf x}^{\prime}, {\bf u}^{\prime},
t^{\prime})$ is known, the bulk of the numerical work in the BGK
scheme lies in solving the above coupled non-linear integral
equations for $g({\bf x}^{\prime}, {\bf u}^{\prime},
t^{\prime})$.

\subsection{Numerical Computation of $f$}
\label{sec-numcompf}

Following the sketch of what is required for a computation of
$f$, we outline some aspects of the numerical procedure, particularly
those which are relevant to adding gravity to the scheme.
The details of the flux computation for the 2D Cartesian case
without gravity are presented in Xu \& Prendergast (1994), 
Xu {\rm et al.} (1996), Kim \& Jameson (1998).
In Cartesian
coordinates we divide the computational domain into cells.  
We then suppose that we
are given the values of mass, momentum
and energy densities within each cell at the beginning of a timestep,
$t_{0}$.  We describe the procedure for computing $f$ at a point
${\bf x}_{f}$ on a cell boundary.

\subsubsection{Numerical Computation of $g$}
\label{sec-numcompg}
We begin the numerical computation of the true distribution function
$f({\bf x}_{f}, {\bf u}, t)$ at ${\bf x}_{f}$ by the construction of 
$g({{\bf x}^{'}},
{{\bf u}^{'}}, t^{'})$ which appears in the integrand of 
(\ref{eq:formalBGKsoltn}).
Because we only need to know $f$ at position ${\bf x}_{f}$ located on a 
cell interface and for the
duration of a CFL timestep we do not
compute 
$g({{\bf x}^{'}}, {{\bf u}^{'}}, t^{'})$ precisely at each point 
along all the
trajectories crossing ${\bf x}_{f}$ at $t$.  Instead
we compute $g$ at the point ${\bf x}_{f}$ on the cell interface  at time $t_{0}$
(where $t_{0}$
corresponds to the beginning of a time step) and then approximate
it in the spatial and temporal 
vicinity of (${{\bf x}}_{f}, t_{0}$) through a Taylor expansion.
\begin{eqnarray}
g({{\bf x}^{\prime}}, {{\bf u}^{\prime}}, t^{\prime})=g({\bf x}_{f}, {\bf u},
t_{0})(1+
\frac{\partial {\rm ln}g}{\partial
{\bf x}^{\prime}} \cdot ({\bf x}^{\prime}-{\bf x}_{f})
\nonumber \\
 + \frac{\partial {\rm ln}g}{\partial
t^{\prime}} \cdot (t^{\prime}-t_{0})+ \ldots )
\label{eq:taylorg}
\end{eqnarray}
To show the velocity dependence in the spatial and time 
derivatives of $g$  we rewrite the Taylor series expansion as:
\begin{eqnarray}
g({{\bf x}^{'}},{{\bf u}^{'}}, t^{'})=g({\bf x}_{f},{\bf u},t_{0})(
1 + {\bf a} \cdot (\bf{x}^{'}-\bf{x}_{f}) 
\nonumber \\
+ \hat{\bf A} \cdot 
(t^{'}-t_{0})).
\label{eq:maxtaylor}
\end{eqnarray}
Here the vectors ${\bf a}$ and $\hat{\bf A}$ depend on the derivatives 
of the parameters of $g$
with respect to space and time and on $\bf{u}$ and $\xi^2$ as well.  
The spatial and time derivatives of $g$ are assumed to
be locally constant and we group them as coefficients of
particle velocities, ${\bf u}^{'}, \xi$ in the following way:
\begin{eqnarray}
{\bf a} & = & {a}_{1} +  {a}_{2} {u}^{'} +  {a}_{3} {v}^{'} + 
{a}_{4} {w}^{'} 
+   {a}_{5} \left( {{\bf u}^{'}}^{2} + {\xi}^{2}
\right) \nonumber \\
\hat{\bf A} & = & \hat{A}_{1}  +  \hat{A}_{2} {u}^{'} + \hat{A}_{3} {v}^{'} +
\hat{A}_{4} {w}^{'}  
+ 
\hat{A}_{5} \left( {{\bf u}^{'}}^{2} + {\xi}^{2}
\right) 
\label{eq:velexp}
\end{eqnarray}
The subscripts $1$, $2$, $3$, $4$, and $5$ refer to the five parameters
in $g$ (cf. eq.~\ref{eq:coupinteqs}).
Recall that ${\bf x}^{'}$ and ${\bf u}^{'}$ 
lie on curved trajectories (eq. \ref{eq:soltnseoms}) if the trajectory
is sufficiently prolonged.
However if we substitute the expressions for ${\bf x}^{'},
{\bf u}^{'}$ given in (\ref{eq:soltnseoms}) into the
expansion for $g({\bf x}^{'},{{\bf u}}^{'},{t}^{'})$
(eq.\ref{eq:maxtaylor}) (with the expressions for ${\bf a}$ and
$\hat{\bf A}$ given in (\ref{eq:velexp}))
 and if we 
retain only terms linear in time, the curvature
terms in the particle trajectories
do not contribute to the expansion.  
Hence to first order in time, the solution of the BGK equation 
for the true
distribution function $f({\bf x}_{f}, {\bf u}, t)$ in one-dimensional Cartesian
coordinates with gravity is equivalent to the solution
for the one dimensional BGK equation in Cartesian coordinates and
without gravity.  This result generalizes to the case of non-Cartesian
geometries.  It is the first important result concerning
the incorporation of gravity into the BGK scheme. \\ 
\indent Our strategy for solving for the terms in the Taylor expansion of 
$g$ is: 
\newline \noindent (a) solve for $g({\bf x}_{f}, {\bf u},
t_{0})$ and the first spatial derivatives
of $g$ from initial conditions. This is possible because as we already
showed in section~\ref{sec-hydfluxesg} there is a direct connection 
between $g$'s
parameters and the mass, momentum and energy densities which are
specified within each cell at the beginning of a timestep $t_{0}$.
We can easily obtain the values of the macroscopic densities and their
first spatial derivatives with respect to ${\bf x}^{\prime}$ at 
$({\bf x}_{f},t_{0})$
through interpolation and thereby solve for $g$ and $g$'s first spatial
derivatives at the cell boundary, ${\bf x}_{f}$. 
We refer to Prendergast \& Xu (1993) for  the details of the numerical
interpolation including a discussion of interpolation switches. The
latter are devices to prevent spurious maxima and other artefacts,
and are commonly used in the reconstruction phase in many schemes.
\newline \noindent (b) before solving for the time derivatives of $g$, we
construct the initial distribution function, $f_{0}$.  Because the
distribution function $f_{0}$ 
should reflect a possible non-equilibrium initial state
across the cell interface located at position ${\bf x}_{f}$, we
assume $f_{0}$ to be composed of two half Maxwellians. It is
easiest to see what we mean by this by again considering a one-dimensional
Cartesian problem.   If we divide  the computational domain into intervals 
whose boundaries are specified
by $x_{1}$,$x_{2}$, $\ldots$, $x_{n}$ and we look at flux transfers
in the $x$-direction through one of these boundaries located at $x_{f}$
then we may construct $f_{0}$ at $x_{f}$ by constructing one half-Maxwellian 
to the left of $x_{f}$ and another half-Maxwellian
to the right of $x_{f}$.
The parameters of the two half Maxwellians as well as their slopes
are obtained from the initial conditions, just as $g$'s parameters
were obtained but the left (right) Maxwellian is computed from
the mass, momentum and energy densities interpolated from the
left (right) side of the cell interface.
This choice for the form of $f_{0}$ is not unique. 
The details
of $f_{0}$ should not matter, since they are rapidly damped for
small $\tau$. 
\newline \noindent (c) Finally we solve for the time derivatives of $g$ by
insisting that the ``compatability condition'' is satisfied on
average over the CFL time interval.  
Due
to our approximate
computation of $g$ and hence of $f$, $g$ and $f$ cannot have exactly the
same moments for the duration of a CFL time step 
over all the trajectories arriving at ${\bf x}_{f}$.
However we insist that they have
the same moments at ${\bf x}_{f}$ and on average over the
time interval $0 \leq t^{'} \leq t$ for which the true distribution function
is computed.
\begin{equation}
\int^t_0 \int {\psi_{\alpha}}^{\rm T} \left( f({\bf x}_{f},{\bf u}, t^{'}) -
g({\bf x}_{f},{\bf u}, t^{'}) \right) d \, \Xi d\, t^{'} = 0
\label{eq:ahats}
\end{equation}
This condition gives an expression for $\hat{\bf A}$ 
(see Xu \& Prendergast (1994)).
It is worth remarking that the manner in which the time
derivatives of $g$ are computed is in some sense implicit.  
Because eq.~\ref{eq:ahats} is applied over the entire time interval,
expressions for $\hat{\bf A}$ arise from information throughout the entire
time interval as opposed to just information 
at the beginning of a time step.  
We also point out that by finding the time-dependence
of $f$ through the formal solution for $f$ plus the 
``compatability condition'' applied on average over
an updating time step, Prendergast and Xu's  (1993) BGK scheme
bypasses the stiffness in the BGK equation which would force the
updating timestep to be the particle collision timescale $\tau$.
Because the updating
time step is determined from the CLF condition hydrodynamics may  be
performed with the BGK scheme. 
Earlier attempts at implementing the BGK equation (e.g. Chu 1965)
were
constrained by the stiffness of the BGK equation and their application
was therefore limited to rarefied gas dynamics.

\subsubsection{The Collision Time, $\tau$}
\label{sec-collisiontime}
Aside from the Courant factor in the computation of the CFL time step
(section~\ref{sec-courant}) (and certain constants which depend
on the adopted interpolation rules), the
expression for the collision time, $\tau$, contains the only 2
parameters in the code.  According to gas kinetic theory, the mean
time between collisions in a gas with density $\rho$ and temperature
$T$ is proportional to $1/(\rho T^{1/2})$.  Since $\lambda \approx 1/T$,
$\tau = {\cal C}_{1} \sqrt{\lambda}/\rho$ where ${\cal C}_{1}$ is a
proportionality constant and the first of the two parameters in $\tau$.
We add a second term to the expression for the collision time. In 
one-dimensional Cartesian coordinates this second term is:
\begin{equation}
{\cal C}_{2}\frac{\arrowvert \sqrt{\lambda_{l}}/\rho_{l} - 
\sqrt{\lambda_{r}}/\rho_{r} \arrowvert}{\sqrt{\lambda_{l}}/\rho_{l} +
\sqrt{\lambda_{r}}/\rho_{r}} \frac{\arrowvert p_{l} - p_{r}  \arrowvert}
{(p_{l}  + p_{r})}.
\label{eq:secondcollterm}
\end{equation}
The subscript $l\hspace{.3em}(r)$ denotes quantities interpolated from the
left (right) side of the cell interface at which we compute
the collision time $\tau$ and $p$ is the gas pressure.  We choose the 
above form for the
second term in the expression for the collision time $\tau$ because
it helps with shocks across which there are gradients in $p$. 
As stated in the introduction (section~\ref{sec-intro}), we motivate 
this second term in the collision time as follows:
given that
the coefficients of heat conduction and viscosity are proportional
to the collision time $\tau$, the second term in the expression for
$\tau$ acts to increase both heat conduction and viscosity at
shock fronts but not at contact discontinuities across which pressure
is continuous.
When the second term in the collision time
dominates,  the order of the scheme is reduced and the true
distribution function $f$ is determined more by the initial
distribution function $f_{0}$ than by the integral over $g$.

\subsection{Computing the Total Fluxes}
Once $f({\bf x}_{f}, t)$ is fully known as a
function of the parameters of $g$, $f_{0}$, the collision
time $\tau$, and $t$, the time-dependent fluxes at ${\bf x}_{f}$
can be computed.  We illustrate flux computation for a one-dimensional 
Cartesian problem.   In that case we are interested
only in the $x$-component of the total flux, i.e. the component of the
flux which is perpendicular to a cell wall at $x_{f}$.  The total amount 
of mass, $x$-momentum and energy densities transferred through the wall
perpendicular to the $x$-direction in a CFL time step
$0 \leq t \leq T$ is found from:
\begin{eqnarray}
\hspace*{-10.cm}\left(\begin{array}{clcr}
{\cal F}^{\rho} \\ {\cal F}^{P_{x}} \\ 
{\cal F}^{{\cal E}_{\rm kin} + {\cal E}_{\rm int}}
\end{array}\right) = 
\nonumber \\
\int^{T}_{0} \int {u_{x}} \left(\begin{array}{clcr}
1 \\ {u_{x}} \\ 
\frac{1}{2}({u_{x}}^{2}+{\xi^2})
\end{array}\right) \, f(x_{f},{\bf u},t) \,
d \, \Xi \, d\, t
\label{eq:fluxes}
\end{eqnarray}
Note that in the end, the gas kinetic origin of
the expressions for the fluxes disappears and they are written as
highly non-linear functions of the mass, momenta and energy densities
which are presumed known at the beginning of the time step.

\subsection{Courant Time Step}
\label{sec-courant}
The time step ${T}$ used to evolve the dynamical equations is chosen so
that it satisfies
the CFL condition.  
Physically the CFL condition limits the distance that information can
travel in one time step to one cell.  
For the one-dimensional
Cartesian case we compute ${T}$ for each cell on the
grid from
\begin{eqnarray*}
{{a}}_{{i+1/2}}
{\hspace*{+.2em}}{T}^{2}   +  \nonumber \\
\left( |{ U}_{i+1/2}|+ c_{i+1/2} \right) {T} +
{{\cal L}}_{i+1/2}   =  0
\end{eqnarray*}
where 
${a}$ is the local gravitational acceleration due to gravity,
{\rm i.e.} $-\frac{\partial \Phi}{\partial {x}}$,
${ U}$ is the local macroscopic velocity, $c$ is the local
adiabatic sound speed and 
\begin{eqnarray*}
{{\cal L}}_{i+1/2} = {\rm min} ((x_{i+1}-x_{i}),
(x_{i}-x_{i-1}),(x_{i-1}-x_{i-2}) \nonumber ).
\end{eqnarray*}
This choice for ${{\cal L}}_{i+1/2}$ insures that flux transport
between adjacent cells is carried out for a length of time short
enough so as not to empty them.

Finally the time step $T$ is chosen from
\begin{equation}
T = (1 - \epsilon) {\rm min} ({T})
\end{equation}
where $(1-\epsilon)$ is a safety factor; it is called the CFL factor
(typically $\epsilon \approx .6$) .

\subsection{Comments about Flux Computation with the BGK method}
\label{sec-implication}

Unlike Riemann solvers which propagate information along
the characteristics of the Euler equations, the BGK scheme 
propagates information along the characteristics of the
Boltzmann equation.  These are particle trajectories in
phase space.  In the BGK scheme
fluid properties along
the full continuum of trajectories passing through $(\bf{x},t)$
contribute to the instantaneous fluxes at $(\bf{x},t)$. 
The
implications of this are many.  
Firstly it bypasses the
limitation of lattice Boltzmann codes which discretize velocity
space thereby restricting the number of directions for the propagation
of information and thereby also limiting the magnitudes of the gas
velocities .

Secondly, using all the trajectories with appropriate weighting 
endows the BGK scheme
with intrinsic upwindedness.  An upwind scheme computes fluxes using
information coming from the same direction as the flow.  When
computing fluxes, the
BGK scheme does not select trajectories corresponding to the direction
of flow and discount the rest;
instead it uses information carried by all
trajectories arriving at the place where we construct $f$, but it
weights the information along each trajectory according to the
number of particles assigned to the trajectory 
by the BGK solution.  This is what we mean
by an intrinsically upwind scheme.  

Thirdly, the consideration of the
full continuum of trajectories for the flux computation creates the potential
for a truly multidimensional code.  Because the velocity
integration includes all propagation directions and because the
trajectories can be followed multidimensionally, the flux in one
direction can potentially be computed from information coming with
all speeds and all directions.  For a genuinely
multidimensional scheme, the interpolation of the initial conditions
for the reconstruction stage must also be multidimensional.  For
example,  
for a computation on a two dimensional Cartesian
grid,  the reconstruction of the initial conditions at a point 
$(x,y)$ should include information from cells surrounding
the point in both the $x$ and the $y$ directions.

At the
present time, the multidimensional BGK schemes which have been
implemented do not take advantage of the truly multidimensional
possibility that the BGK scheme offers.  They still simplify flux
updates by separating them into individual operations in 
independent directions.  This is an example of
``operator-splitting''. On the optimistic side, a truly
multidimensional BGK scheme is conceivable.  The same cannot be said
for schemes based on Riemann solvers.  Even in two dimensions it is
not feasible to construct a truly multidimensional Riemann solver. 
Upon representing the initial
conditions as piecewise constant in two-dimensional
cells, the Riemann solution is straightforward to apply at each of the 
interfaces of a two-dimensional cell but problematic at cell corners.
Furthermore, by propagating information  only along the directions
normal to cell interfaces, a two-dimensional Riemann solver has
the same limitation as an operator-split method: it does not accommodate
the possibility that waves in two-dimensional flows may propagate in
infinitely many directions (Roe 1986).

\subsection{The Source Terms}
\label{sec-src}
Because to first order in time the computation of the true
distribution function, $f({x},\bf{u},t)$, and hence the
computation of the hydrodynamical fluxes is unaffected
by gravity, gravity's influence on the gas flow is relegated
to source terms.  

\subsubsection{Gravitational Momentum Source Terms}
\label{sec-gmomsrc}
In one-dimensional Cartesian coordinates gravitational source terms 
contribute to the momentum change in a cell 
over a CFL time step $t=0$ to $t=T$ in the following form: 
\begin{equation}
{\cal S}_{P_{x}}(x_{i-1/2},T)=-\int^T_0 \int^{x_{i}}_{x_{i-1}}
\rho(x^{\prime}, t^{\prime})
\frac{\partial \Phi}{\partial x}(x^{\prime}, t^{\prime}) dx^{\prime}dt^{\prime}.
\label{eq:momsrcterm}
\end{equation}
\noindent This integral may be computed in a number of ways. 
One possible approach is to do a linear interpolation in $x^{\prime}$
using the values of the  
mass densities and forces at the cell wells
surrounding the cell for which we wish to compute the source term.

The time integration required by (\ref{eq:momsrcterm}) takes advantage
of the fact that we know the new mass densities at the end of an 
updating time step, {\rm i.e.} at $t=T$, before we've completed the
updates of the other fluid quantities, {\rm i.e.} $P_{x}$,
${\cal E}_{\rm kin}+{\cal E}_{\rm int}+{\cal E}_{\rm grav}$.  
This is a consequence of there being no source terms in the
mass continuity equation.
\noindent From the mass densities at times $t=0$
and $t=T$ we may perform a bilinear interpolation 
in $x$ and $t$ for the 
integral in equation (\ref{eq:momsrcterm}). Specifically 
\begin{eqnarray}
\hspace*{-200em}{\cal S}_{P_{x}}(x_{i-1/2},T) &  =  & \nonumber \\
-\frac{T}{2}
\int^{x_{i}}_{x_{i-1}} \{
\rho(x_{i-1},0)\frac{\partial \Phi}{\partial x}(x_{i-1},0)
(1-\beta(x^{\prime}))  &  +  &
\nonumber \\
\rho(x_{i},0)
\frac{\partial \Phi}{\partial x}(x_{i},0)
(\beta(x^{\prime})) &  +  & 
\nonumber  \\
\rho(x_{i-1},T)
\frac{\partial \Phi}{\partial x}(x_{i-1},T)
(1-\beta(x^{\prime}))  &  +  & 
\nonumber \\
\rho(x_{i},T)
\frac{\partial \Phi}{\partial x}(x_{i},T)
(\beta(x^{\prime}))  \} 
 dx^{\prime}  
\label{eq:gmomsrc}
\end{eqnarray}
where 
\begin{displaymath}
\beta(x^{\prime})=\frac{(x^{\prime}-x_{i})}{(x_{i+1}-x_{i})}.
\end{displaymath}

\subsubsection{The Energy Equation}
\label{sec-energy}
As for gravity's contribution to the energy equation,
we explore two versions of it.
One form of the energy equation: 
\begin{displaymath}
\frac{\partial}{\partial t} \left( {\cal E}_{\rm kin}+{\cal E}_{\rm int}\right)
+ \frac{\partial}{\partial x} \left( {
F}^{{\cal E}_{\rm kin}+{\cal E}_{\rm int}} \right) -
P_{x} \frac{\partial \Phi}{\partial x} =0 \nonumber
\end{displaymath}
has a gravitational energy source term, $
- P_{x} \frac{\partial \Phi}{\partial x}$.  The gravitational 
energy source term may be computed 
analogously to
the way in which the gravitational momentum source terms are computed
({\rm cf.} eq. \ref{eq:gmomsrc}).

As mentioned in section~\ref{sec-hydeq}, another form of the energy
equation incorporates the gravitational
energy source term into the energy fluxes and hence is in conservative
form:
\begin{displaymath}
\frac{\partial}{\partial t} \left( {\cal E}_{\rm kin}+{\cal E}_{\rm int}+{\cal E}_{\rm grav}\right)
+ \frac{\partial}{\partial x}
\left( { F}^{{\cal E}_{\rm kin}+{\cal E}_{\rm int}} +
\Phi {{F}}^{\rho} \right) =0. \nonumber
\end{displaymath}
The conservative form of the energy equation requires a good
computation of the gravitational energy because now it, in addition
to the kinetic energy, ${\cal E}_{\rm kin}$, must be subtracted from
the total energy for the purpose of obtaining the internal energy,
${\cal E}_{\rm int}$. These subtractions are potentially 
extremely delicate.  They are important because the internal energy is used in computing $\lambda$ (which is inversely proportional to the temperature),
a crucial parameter for the hydrodynamic fluxes.

Because the form of the gravitational energy is similar to the
form of the gravitational source terms,
\begin{equation}
{\cal E}_{\rm grav}(x_{i-1/2})=\int^{x_{i}}_{x_{i-1}}
\rho(x^{\prime})
\Phi(x^{\prime}) dx^{\prime}
\end{equation}
it may be numerically computed in the same manner, {\rm i.e.} by a linear
interpolation to the values of the densities and the potentials
at the walls bounding a cell.  

For the gravitational flux term, 
$\frac{\partial}{\partial x}(\Phi {{
F}}^{\rho})$, in the conservative energy equation
we use the value of the gravitational potential $\Phi$ at 
the same point $x_{f}$ at which we construct the true distribution 
function, $f$, for the
other  fluxes.  

\subsection{Summary of the BGK Method with Gravity}
We have given a  description of the procedure
for computing hydrodynamical fluxes and source terms in a BGK scheme
designed for hydrodynamics in the
presence of a time-independent gravitational field.

\noindent 1.)While incorporating gravity into the BGK scheme we have found
that to first order in time gravity does not alter the computation
of the distribution function from which hydrodynamical
fluxes are computed.  It enters into the hydrodynamical computation
through gravitational momentum source terms as well as through either a
gravitational energy source term or through gravitational energy
fluxes.  The latter depends on the chosen form of the energy equation.

\noindent 2.)The energy source term formulation by definition does 
not guarantee conservation
of total (kinetic + internal + gravitational) energy.  An inaccurate
computation of the energy source term introduces numerical heating
and/or cooling. Because the energy equation can be manipulated into
a conservative form for the total kinetic + internal + gravitational
energy, computation of an energy source term is avoidable.
In lieu of an energy source term, the gravitational energy enters 
into the hydrodynamical
computation through flux terms.  These flux terms do not require any
flux computation in addition to the computation which is already required
in a hydrodynamics scheme without gravity.  This is because they are 
conveniently the product of the gravitational potential and the mass 
flux terms.  

\noindent 3.)Even though an implementation of
the conservative form of the energy equation by definition 
assures total energy conservation, it introduces another challenge
brought on by the inclusion of gravity into a hydrocode.  For the
calculation of hydrodynamical fluxes in the BGK scheme it is 
essential to accurately compute the temperature
and therefore the internal energy.  With gravity and with an 
implementation of the conservative form of the energy equation 
it therefore becomes necessary to accurately compute the gravitational energy 
so that it, in addition to the kinetic energy, may be subtracted from 
the total energy (kinetic + internal + gravitational) 
to give the internal energy without introducing numerical heating (or
cooling) into the scheme.

\section{Results}
\label{sec-results}
We present results from two tests of the BGK scheme in the
presence of a time-independent gravitational potential.  
The first test
reveals the effect on a simulation of using different forms
of the energy equation.  The second test is performed on
an axisymmetric cylindrical grid.  Both tests reveal the
capability of the BGK scheme with gravity to reach an equilibrium
state and to maintain a gas configuration in hydrostatic equilibrium.

\subsection{On a 1D Cartesian Grid: Gas Falling into a Fixed External Potential}
\label{sec-ho}
We perform this test case to compare the results from two versions of
the BGK scheme with gravity: one using the conservative form of the
energy equation (the ECS scheme) and one using the energy source term 
formulation of the energy equation (the EST scheme). 
The initial conditions for the two simulations are identical. Each 
simulation is performed on a 1D Cartesian grid with 68 evenly spaced cells.  
The gas is initially stationary ($P_{x}=0$, where $P_{x}$ is the $x$-momentum) 
and homogeneous with mass density
$\rho = 1$ and internal energy density $\cal{E}_{\rm{int}}={\rm 1}$
in each cell.  The
external gravitational field is constant in time and its potential, $\Phi$ has
the form of a
sine wave, 
{\rm i.e.} $\Phi = - \Phi_{0}({\cal L}/(2\pi)) {\rm sin}(2\pi x/{\cal L})$ 
with $\Phi_{0} = 0.02$
and ${\cal L} = 64$.  Because we apply periodic boundary conditions 
there are two ghost cells at each end of the grid. We take $\epsilon$
in the Courant condition to be $.4$, $\gamma = \frac{5}{3}$ and 
${\cal{C_{{\rm 1}}}}{= \rm .01}$
and ${\cal{C_{{\rm 2}}}} {\rm = 1}$ in the expression for the collision time
$\tau$.  We run both versions of the code for $500,000$ iterations.
The total time at the end of the run with the ECS scheme is
$285739.54$ in machine units ($4706$ in units of the initial sound
crossing time).  Note that we define the total time to be the sum of
the lengths of each of the individual CFL timesteps used for the
updates.  For the run with the EST scheme
the total time at the end of the run is $261888.6$ in machine
units ($4313$ in units of the initial sound
crossing time).

\begin{figure*}
\epsfbox{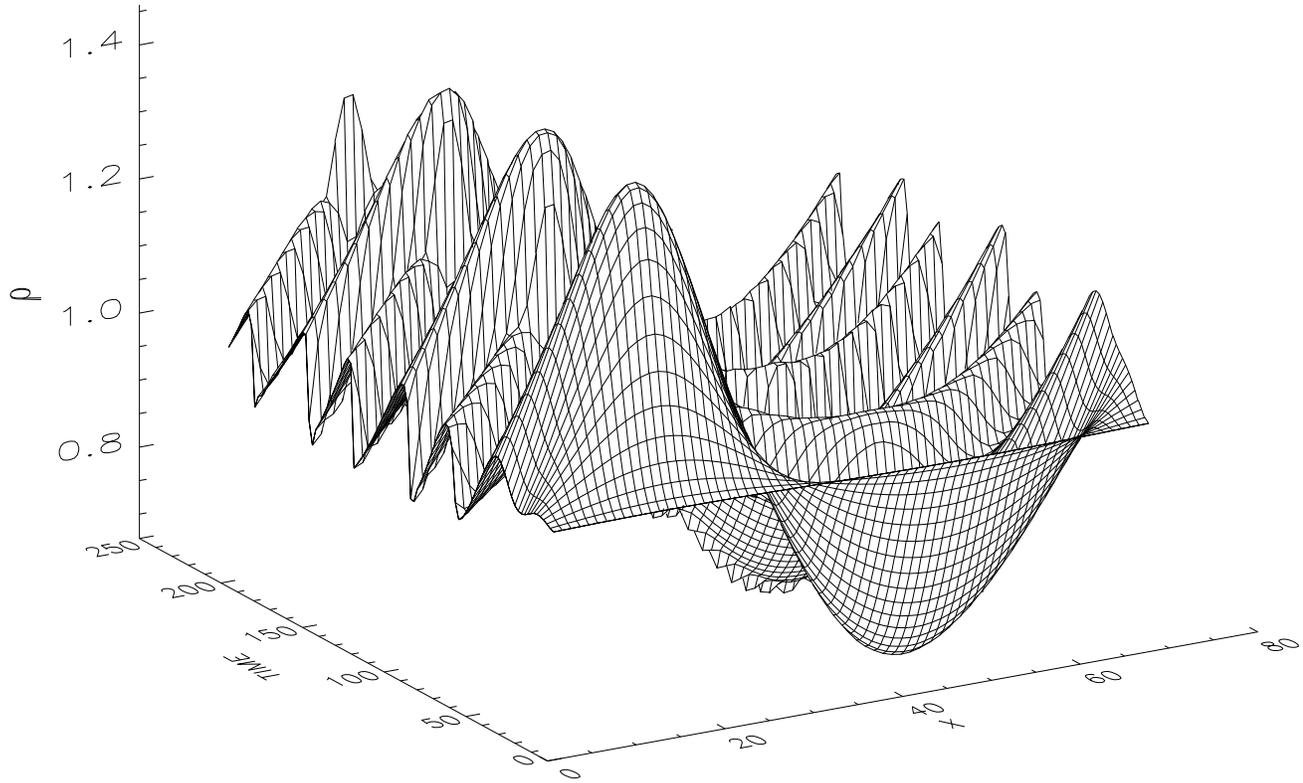}
  \caption[]{Early stage of time evolution of mass density profile
for run with total energy conserving (ECS) scheme, including
gravitational energy. Total time 201.19 ($\equiv 3.3$ 
sound crossing times).
}
  \label{ecsshortdens}
\end{figure*}

\begin{figure*}
\epsfbox{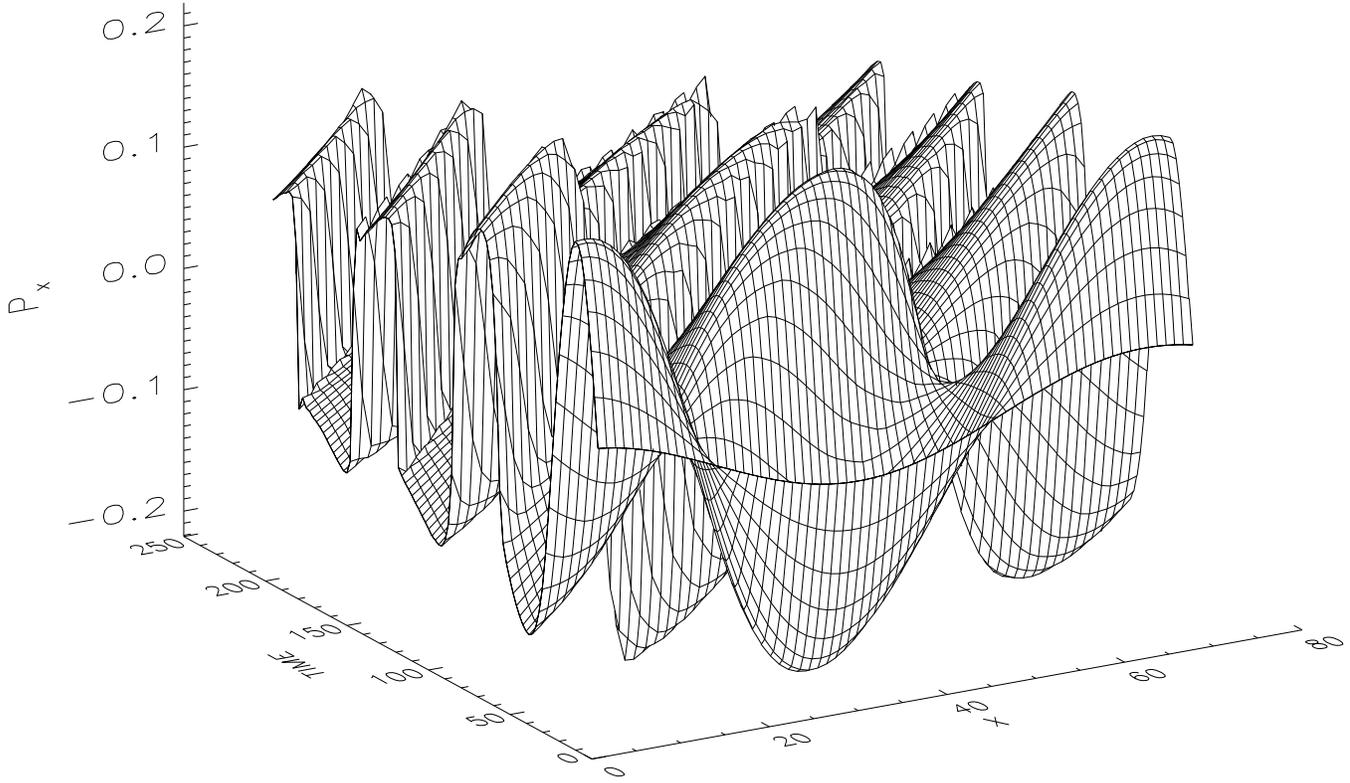}
  \caption[]{Early stage of time evolution of $x$-momentum density profile
for run with ECS scheme. Total time 201.19 ($\equiv 3.3$ 
sound crossing times).
}
  \label{ecsshortxmom}
\end{figure*}

\begin{figure*}
\epsfbox{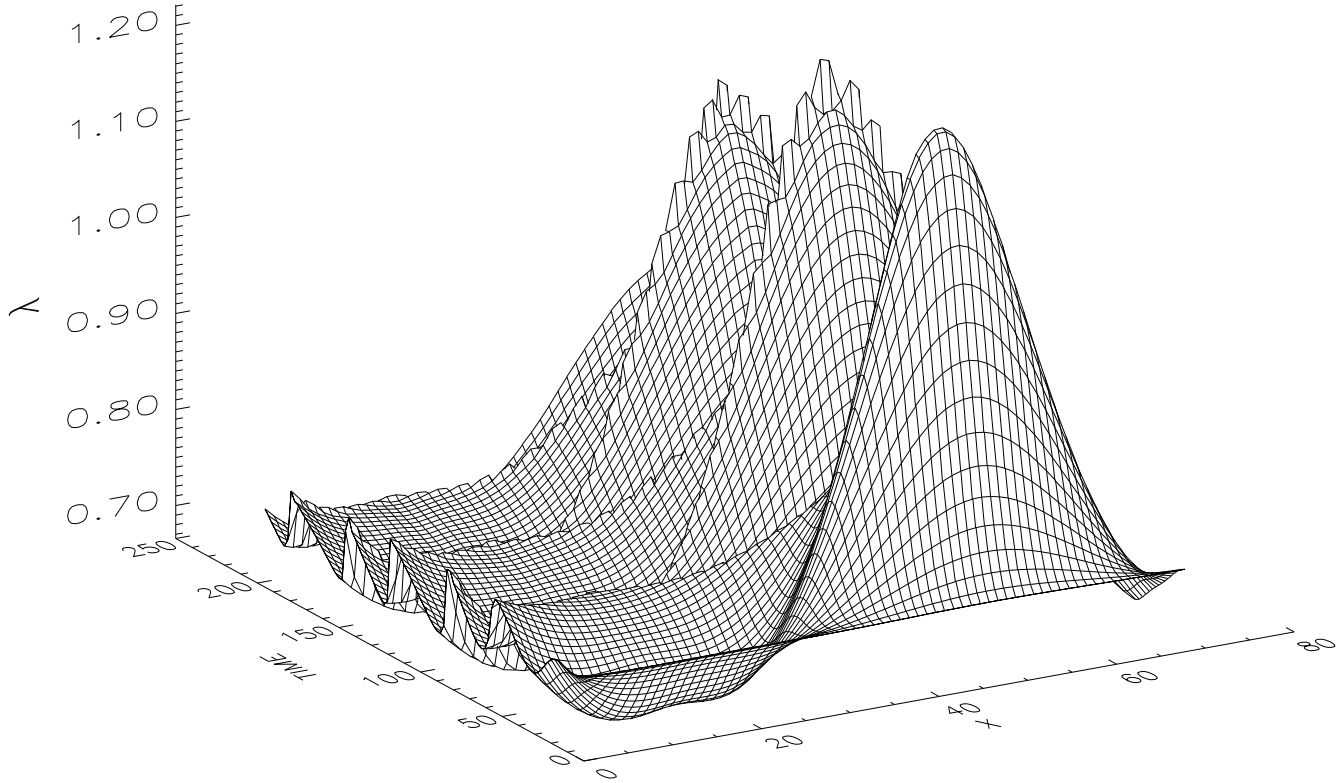}
  \caption[]{Early stage of time evolution of $\lambda$ profile
for run with ECS scheme. Total time 201.19 ($\equiv 3.3$ 
sound crossing times).
}
  \label{ecsshortlambda}
\end{figure*}

\begin{figure*}
{\epsfxsize=18.cm \epsfysize=11.cm \epsfbox{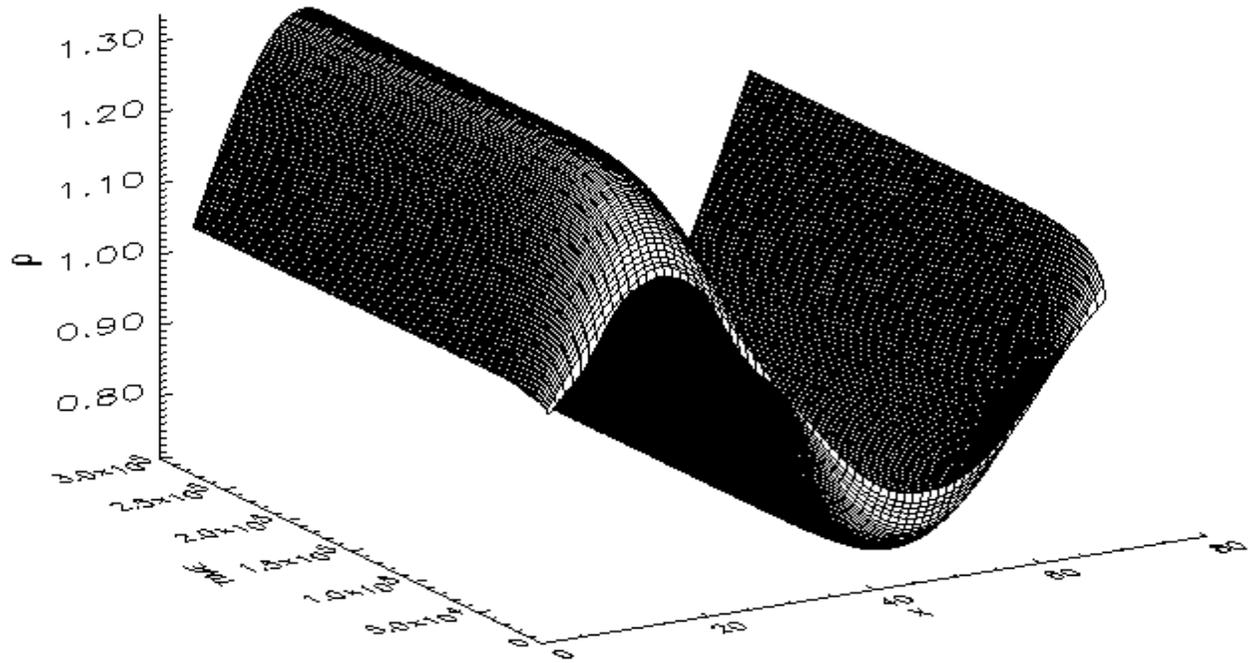}}
  \caption[]{Mass density profile for long run with ECS scheme after
total time of 285739.54 ($\equiv 4706$ 
sound crossing times).
}
  \label{ecsdenslong}
\end{figure*}

\begin{figure*}
{\epsfxsize=18.cm \epsfysize=11.cm \epsfbox{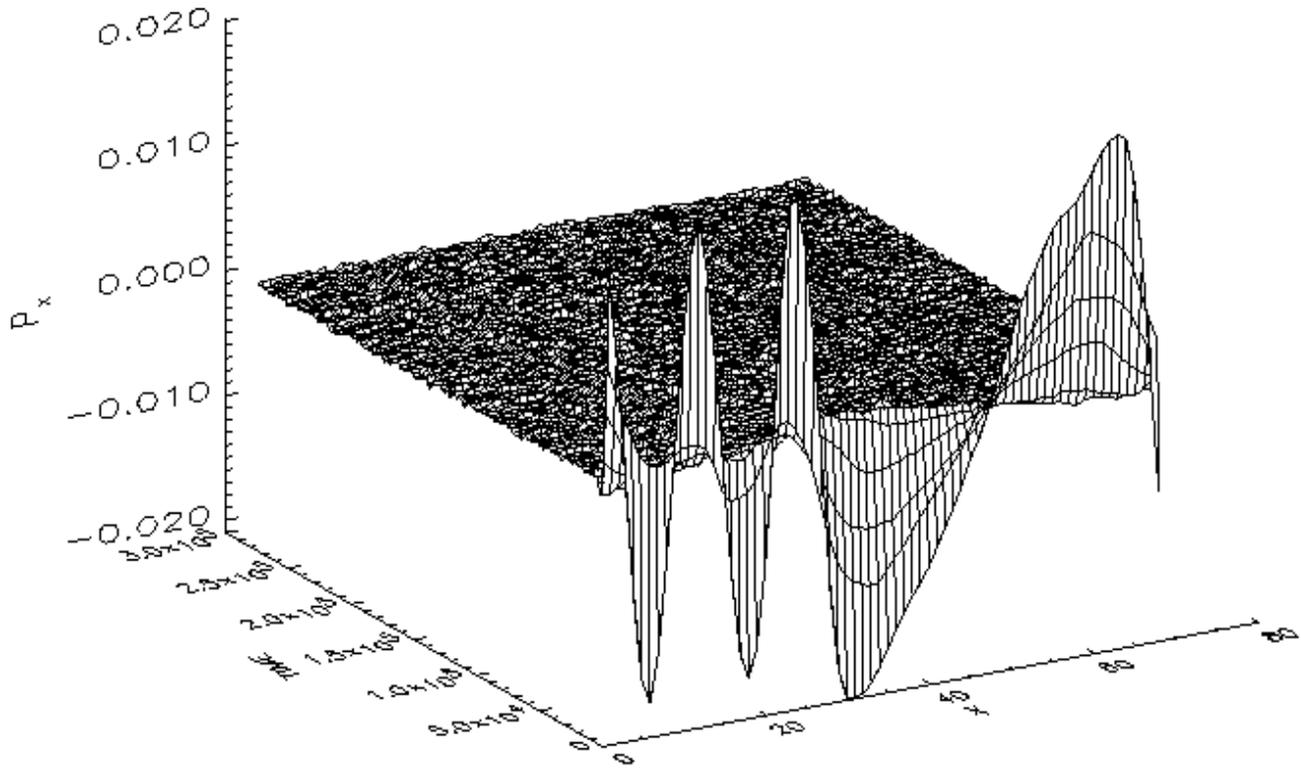}}
  \caption[]{$X$-momentum density profile for long run with ECS scheme after
total time of 285739.54 ($\equiv 4706$ 
sound crossing times).
}
  \label{ecsxmomlong}
\end{figure*}

\begin{figure*}
{\epsfxsize=18.cm \epsfysize=11.cm \epsfbox{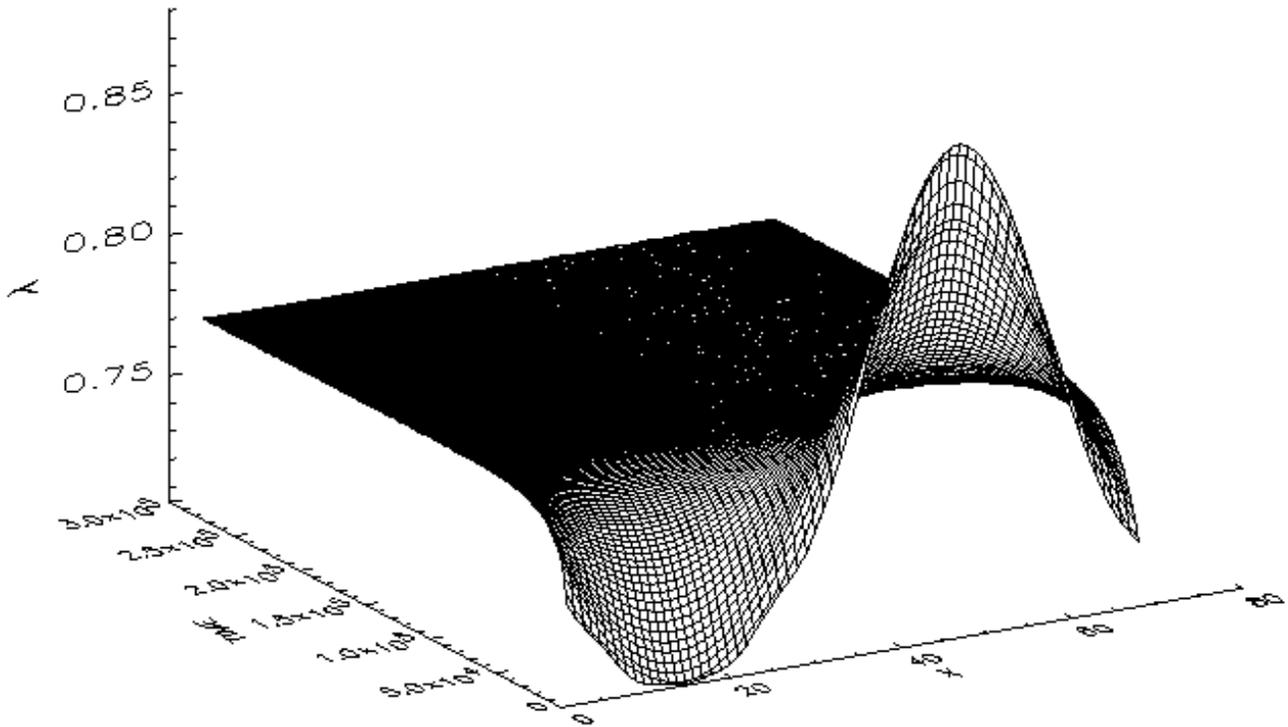}}
  \caption[]{Lambda profile for long run with ECS scheme after
total time of 285739.54 ($\equiv 4706$ 
sound crossing times).  Note that lambda becomes and remains constant
after some time.  This is entirely in accord with expectation of the 
long-term behaviour of a closed system: the temperature (i.e. $\frac{1}{\lambda})$ reaches equilibrium and stays there.  Compare this result with
fig.~\ref{estlambdalong} which shows long term heating of numerical origin
using a different version of the code.}
  \label{ecslambdalong}
\end{figure*}

\begin{figure*}
{\epsfxsize=18.cm \epsfysize=11.cm \epsfbox{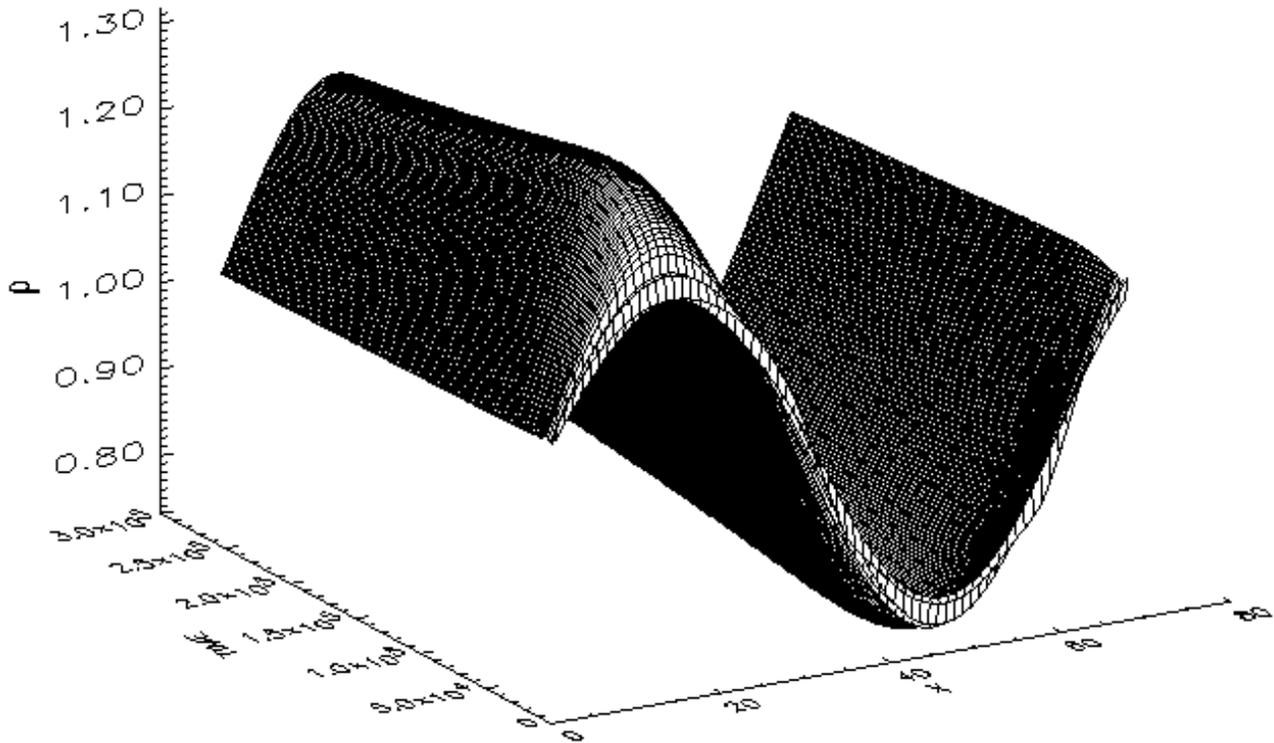}}
  \caption{Mass density profile for long run with energy source term 
(EST) scheme after
total time of 261888.6 ($\equiv 4313$ 
sound crossing times).  Note that the density profile has not settled to
a constant profile with time.  Compare with fig.~\ref{ecsdenslong} 
which uses the ECS scheme.}
  \label{estdenslong}
\end{figure*}

\begin{figure*}
{\epsfxsize=18.cm \epsfysize=11.cm \epsfbox{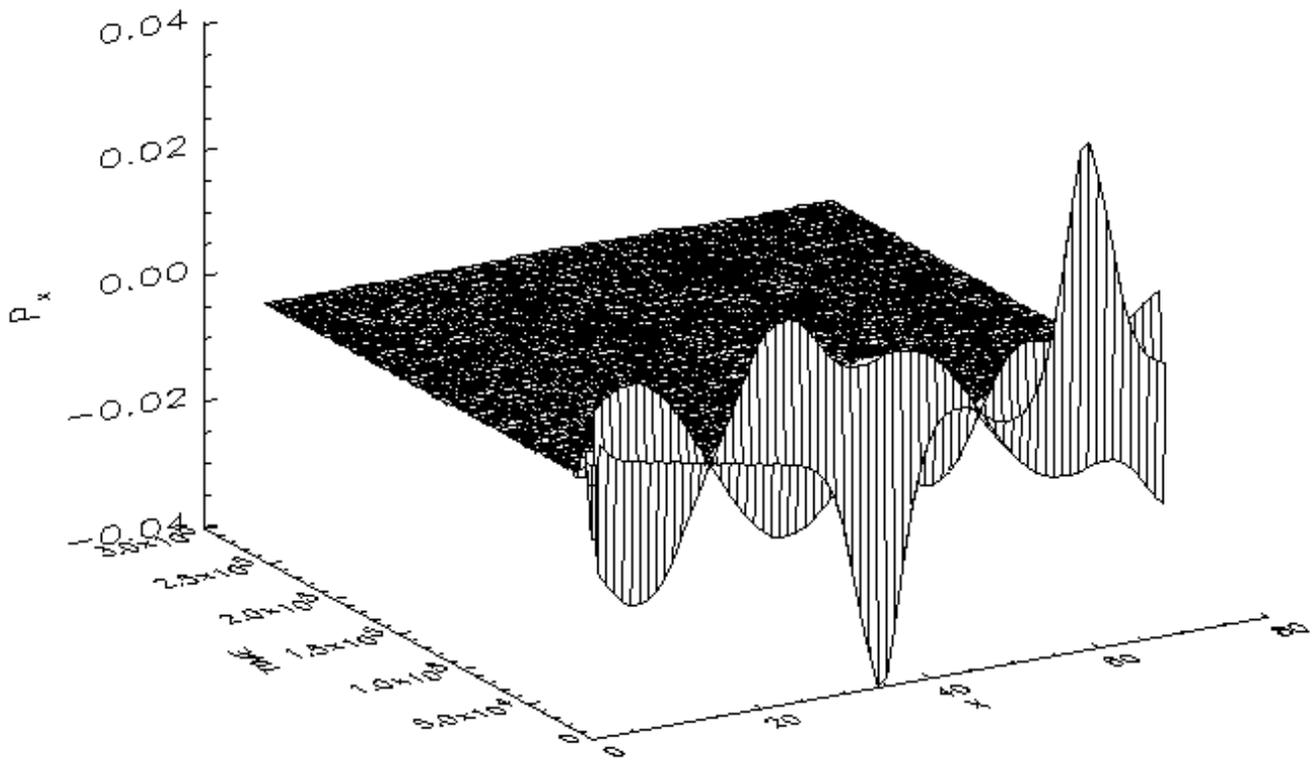}}
  \caption[]{$X$-momentum density profile for long run with EST scheme after
total time of 261888.6 ($\equiv 4313$ 
sound crossing times).
}
  \label{estxmomlong}
\end{figure*}

\begin{figure*}
{\epsfxsize=18.cm \epsfysize=11.cm \epsfbox{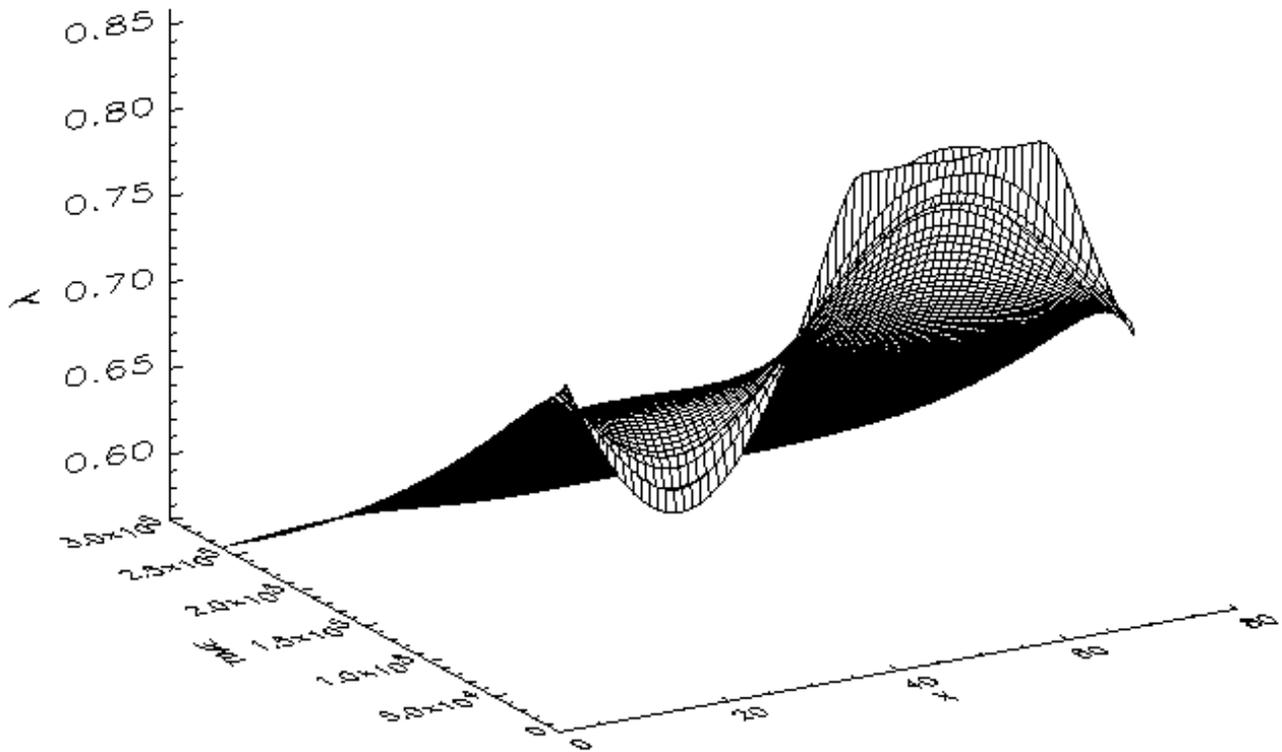}}
  \caption[]{Lambda profile for long run with EST scheme after
total time of 261888.6 ($\equiv 4313$ 
sound crossing times).
Note that lambda descreases instead of remaining constant (as it should)
at large time.  The effect is due to numerical heating and is absent in the
results from the totally conservative (ECS) scheme 
(cf. fig.~\ref{ecslambdalong}).}
  \label{estlambdalong}
\end{figure*}

Figs.~\ref{ecsshortdens} through~\ref{ecslambdalong} show results 
from the version of the
BGK scheme which implements the energy conserving form of the energy
equation and which therefore keeps track of the total kinetic +
internal + gravitational energy density within each cell.  
Figs.~\ref{estdenslong} through~\ref{estlambdalong} on the other
hand show results from an implementation of the energy equation in
a non-conservative form, {\rm i.e.} one in which gravity is incorporated
into the energy equation through a gravitational source term. 
The plots show the time evolution of various quantities ({\rm i.e.} mass
density, $x$-momentum and lambda) as
functions of  position on the grid.  Because the cells are evenly
spaced with a spacing of $\Delta x = 1$, we label cell positions by
an index X which runs from $1$ to $68$ in the plots. The quantity on
the vertical axis in each of the plots is also plotted at approximately
equal time intervals.

For the simulations using the energy conserving scheme we present
not only plots showing the long term evolution of the gas, but also
some (Figs.~\ref{ecsshortdens},~\ref{ecsshortxmom},~\ref{ecsshortlambda}) 
which show the evolution of the gas early on in the simulation.
In these figures, particularly in the $x$-momentum plot 
({\rm i.e.} Fig.~\ref{ecsshortxmom}), we see the steepening of the sound
waves into shocks.
With the
values for the internal energy density and the mass density given in the
initial conditions, the initial sound crossing time for the gas is
$60.72$ (in machine units).  For the data presented in 
figures~\ref{ecsshortdens},~\ref{ecsshortxmom} and~\ref{ecsshortlambda} 
the total
time elapsed is $201.19$ (in machine units). This is
equivalent to $3.3$ sound crossing times and indeed we see that the
gas has undergone a bit more than three oscillations in these figures.

In the plots showing the long term evolution of
the gas (Figs.~\ref{ecsdenslong} to~\ref{estlambdalong}),
at the time of the earliest curve (time $\approx 2816$) 
the gas has experienced about $46$ sound crossing times (we compute this
using the value for the initial sound crossing time).
The mass density in both the ECS scheme and the 
EST scheme  at $t= 2816$ is high at the location
of the gravitational potential minumum and low at the location
of the potential maximum indicating that the gas has fallen into the
potential well.  Because the gas should convert some of its
gravitational energy into internal energy as it falls into 
the potential well, we look
to see if the gas at the density maximum is hotter.
As expected, the temperature at $t= 2816$
is highest (hence lambda is lowest) 
at the location of maximum density and the temperature is lowest
(hence lambda is highest) at the location of minimum
density.  For reference, lambda at the beginning of the simulations
is $0.75$.

In the results from the scheme using
the non-conservative form of the energy equation, there is a monotonic 
decline in lambda (Fig.~\ref{estlambdalong}) at increasing times.  Since 
lambda is inversely
proportional to the temperature, this decline corresponds to a
temperature rise which we believe is due to ``numerical heating''.
This heating persists with time and therefore deters
the gas from reaching thermodynamic equilibrium.  The heating is
reflected in the value of the total energy (kinetic + internal +
gravitational) summed over the entire grid at the end of the run:
it is $~\approx 10 \% $ greater than the initial total energy.

\begin{figure*}
\rotate[r]{\epsfbox{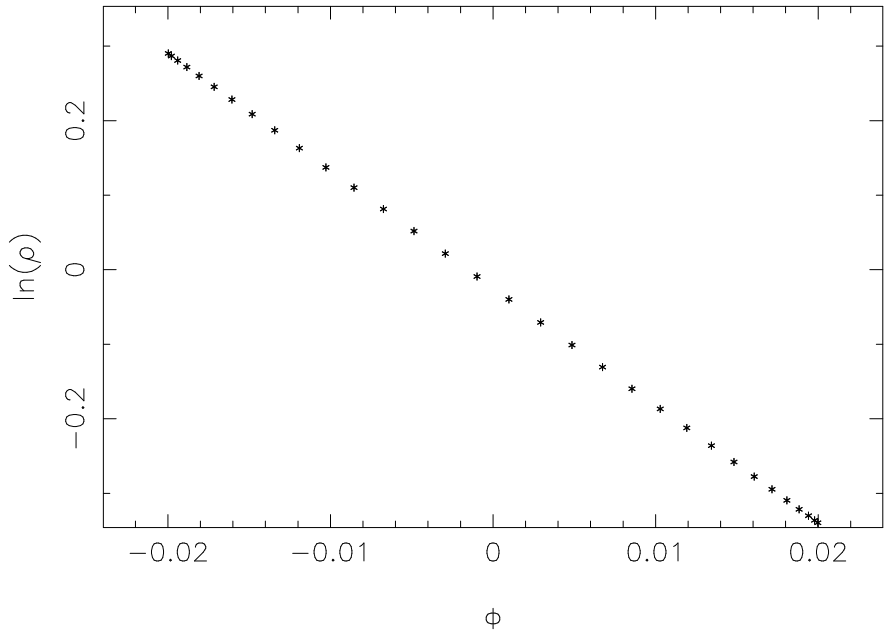}}
  \caption[]{Check to see how well final density profile from long run
with ECS scheme matches the
analytic solution for the equilibrium profile.  The points should fall
on a straight line if there is a perfect match to the analytic solution.
See text (section~\ref{sec-ho}) for parameters of a least squares fit to the
points.}
  \label{ecsequ}
\end{figure*}

In contrast,
the results from the run with the energy conserving implementation
of the energy equation show a gas which reaches thermodynamic
equilibrium: the lambda plot (Fig.~\ref{ecslambdalong}) is level at late 
times with a mean value of $.77$ (deviation $\approx 1.52D-04$) 
and remains at the same value throughout the entire grid
for as long as we run the simulation.  
The difference between the initial and final total energies on the
grid is ${\rm 4.3451D-13}$.
In addition to thermodynamic equilibrium, the gas reaches 
mechanical equilibrium as verified by the
$x$-momentum plot: at the final time plotted $P_{x}$ is
$-1.16D-20$ in the mean with a deviation of $2.04D-04$.
We believe that the noisiness in the final $P_x$ state is a consequence of 
the run being performed without the use of interpolation switches for the
flux computation.  
The analytic solution for the equilibrium mass density profile in a fixed
gravitational potential is
$\rho(x) = \rho(0) e^{-(\frac{\mu m_{H}}{ k T}) \Phi }$ 
where $\mu$ is the atomic weight, $m_{H}$ is the
mass of a hydrogen atom, $k$ is Boltzmann's constant, $T$ is the
temperature, $\Phi$ is the gravitational potential.  
Because the temperature
should be constant throughout the grid at equilibrium, a plot of 
the numerical values of ${\rm ln}(\rho)$
vs. $\Phi$ displays how closely the gas has settled to equilibrium.
We plot these quantities for the results from the ECS scheme in
Fig.~\ref{ecsequ}.  When we fit a line ${\rm ln}(\rho)_{i} =
a + b \Phi_{i}$ to the numerical results then $a=-2.46484E-02$, 
$b=-15.7485$ and the variance of $a$ is $2.47615E-05$ while 
the variance of $b$ is $1.75090E-03$.

\subsection{On an Axisymmetric Grid: Gas Falling into a Fixed External Potential}

\begin{figure*}
\rotate[l]{\epsfxsize=15.cm \epsfysize=17.cm \epsfbox{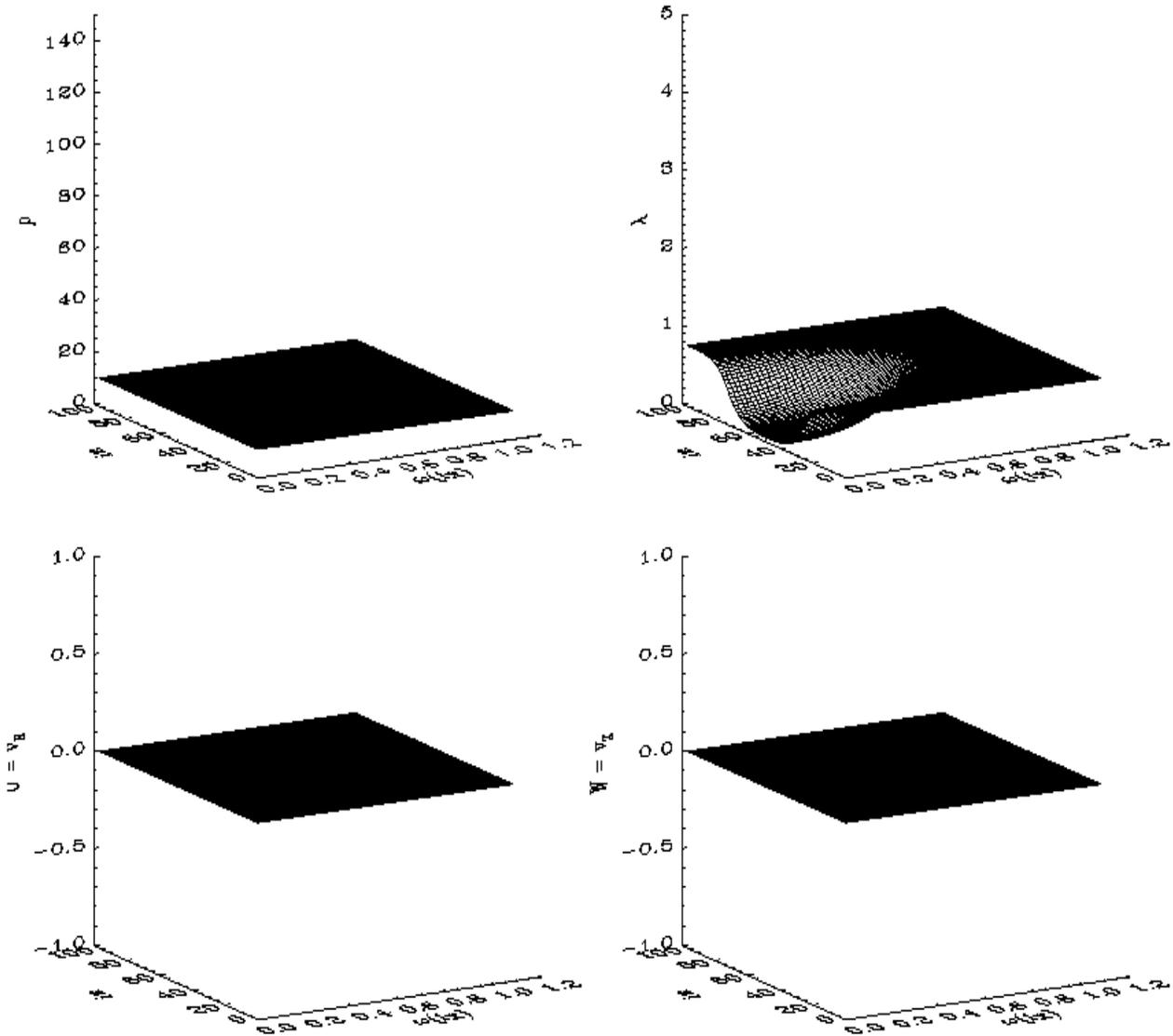}}
  \caption[]{The initial conditions for the simulation of gas
falling into a fixed external spherically symmetric gravitational
potential well: uniform density (top left), lambda (top right), zero
radial velocity (bottom left) and zero vertical velocity (bottom right). } 
  \label{thinitfixpot}
\end{figure*}

\begin{figure*}
\rotate[l]{\epsfxsize=15.cm \epsfysize=17.cm \epsfbox{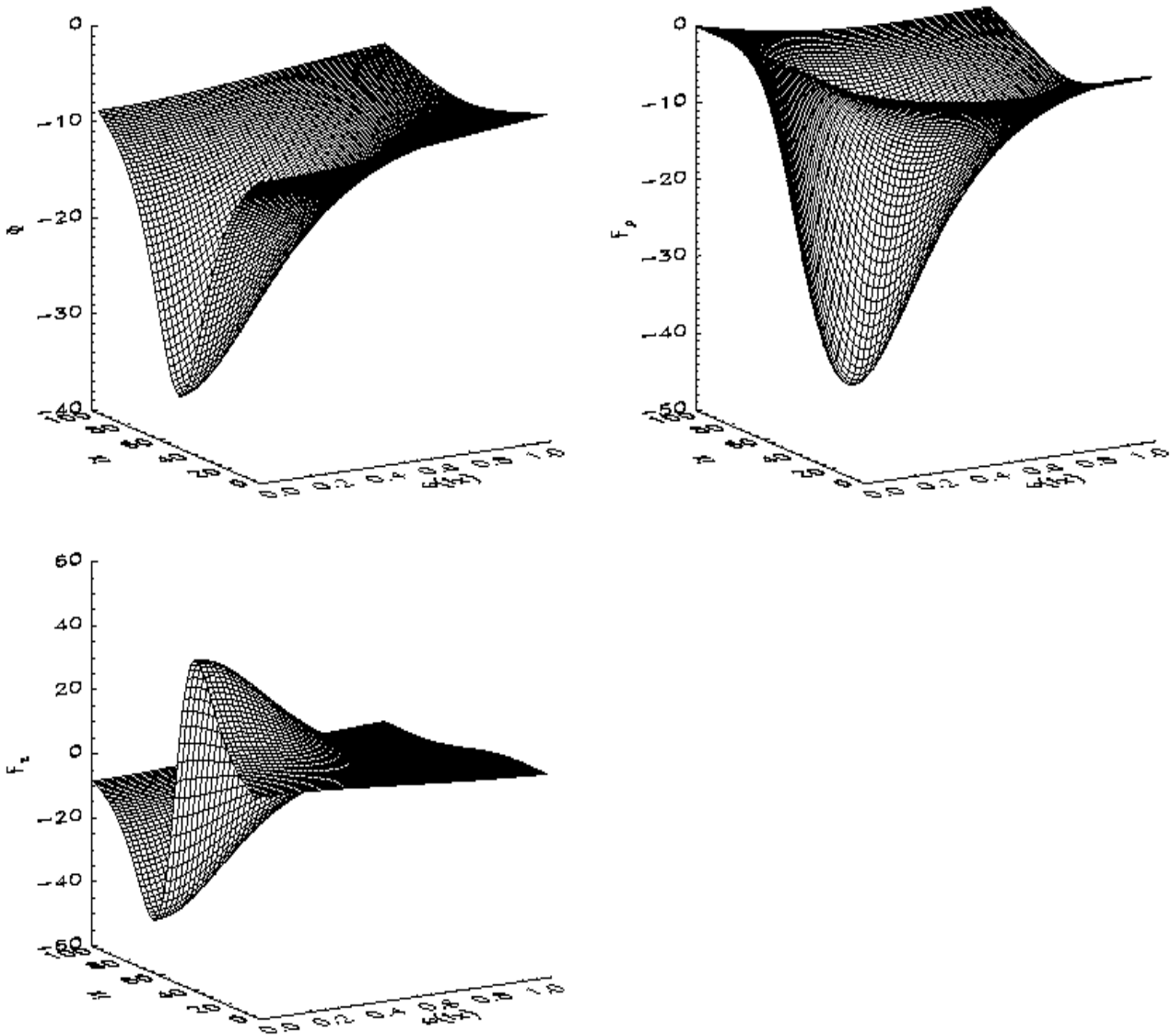}}
  \caption[]{The gravitational potential (top left),
the radial force (top right) and the vertical force (bottom left), 
each as a function of
$\varpi$ and $z$.  They are fixed in time and space throughout the
entire run.  } 
  \label{fixgforce}
\end{figure*}

We simulate the shock heating of gas falling into a
fixed, external gravitational potential well in 2D. The hydrodynamic 
computation is performed on an axisymmetric
cylindrical grid which has $50$ logarithmically
spaced cells in radius, $\varpi$, and $100$ evenly spaced cells in the
$z$ direction.  
The gas falls into a gravitational 
potential well (Fig.~\ref{fixgforce}) which is
derived from a spherical density distribution with the following 
profile: $\rho \sim (1 + \frac{r^{2}}{{r_{\rm max}}^{2}})^{-2.5}$.
The potential is centered on the origin of the axisymmetric
cylindrical grid, namely at $\varpi=0$ and at the midplane in $z$.
The boundary conditions for the simulation are reflecting.
The gas is initially stationary and uniform
in density (Fig.~\ref{thinitfixpot}).  As seen in
Fig.~\ref{thinitfixpot}, it is initially higher in temperature
at the location of the potential well than in other parts of the
grid. In machine units, 
$G=7$, $\rho=10$, the radial momentum and vertical momenta densities
in each cell are $0$, {\rm i.e.} $P_{\varpi}=0$ and $P_{z}=0$.  The internal
energy density in each cell is
initially set to ${\cal E}_{\rm int} = 2.5 {\rm X} 10^{-14} - {\cal
E}_{\rm grav}$, where ${\cal E}_{\rm grav}$ is the gravitational
energy density in each cell.  The axisymmetric cylindrical grid
extends from $-1.1$ to $+1.1$ in the vertical direction and from
$0$ to ${\rm r_{max}} = +1.1$ in the $\varpi$ direction.  We also take $\epsilon$
in the Courant condition to be $.8$, $\gamma = \frac{5}{3}$,  and 
${\cal{C_{{\rm 1}}}}{= \rm .001}$
and ${\cal{C_{{\rm 2}}}} {\rm = 1}$ in the expression for the collision time
$\tau$.

\begin{figure*}
\epsfxsize=18.0cm \epsfysize=21.cm \epsfbox{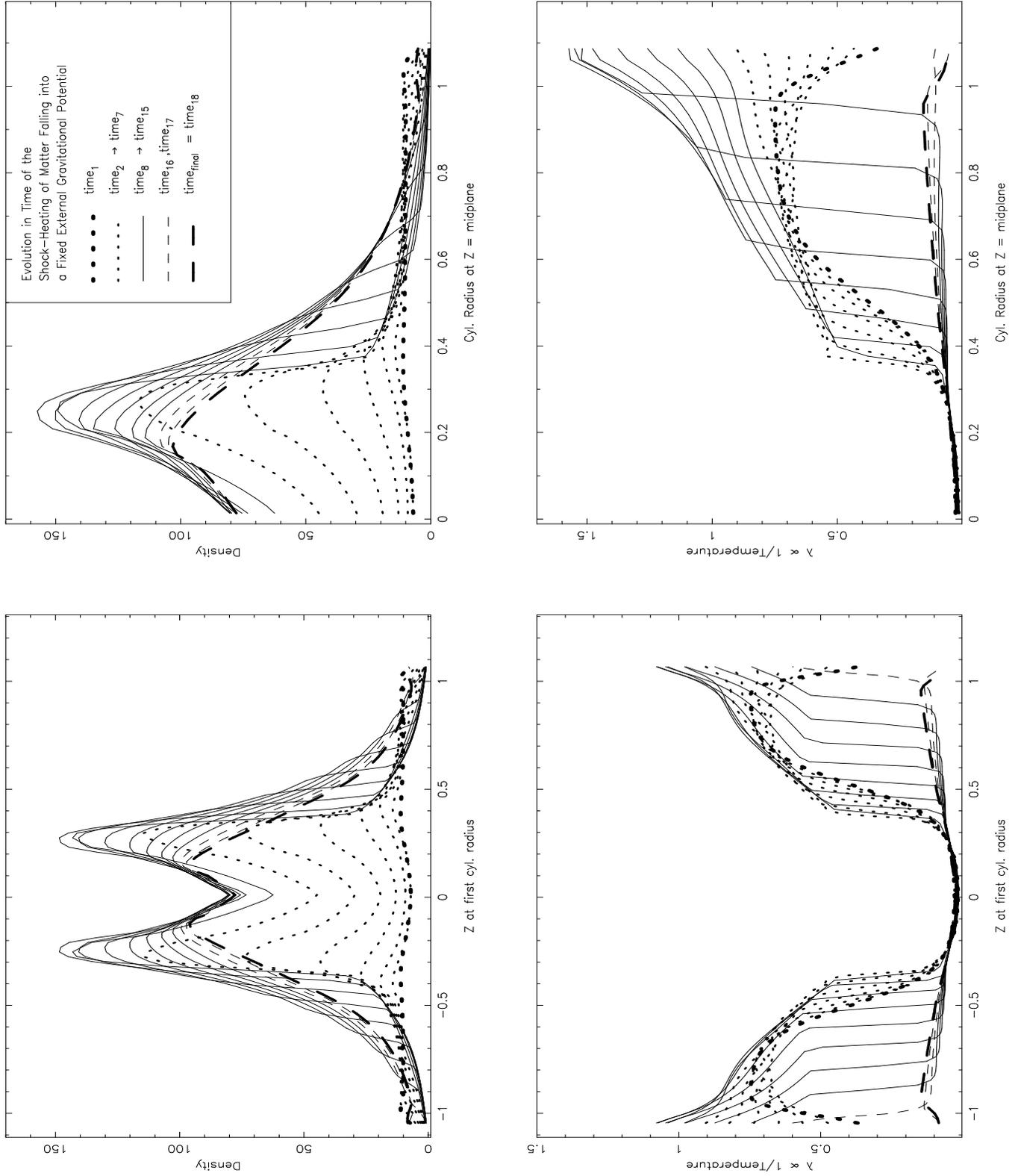}
  \caption[]{Results from simulations of the shock heating of gas
falling into a fixed
  external spherically symmetric gravitational potential well.
Time evolution of the density (top) and lambda (bottom) along
two  rays of the cylindrical grid. The cylindrical grid has
50 cells in $\varpi$ and 100 cells in $z$.} 
  \label{crossectdenslam}
\end{figure*}

\begin{figure*}
\epsfxsize=18.0cm \epsfysize=21.cm \epsfbox{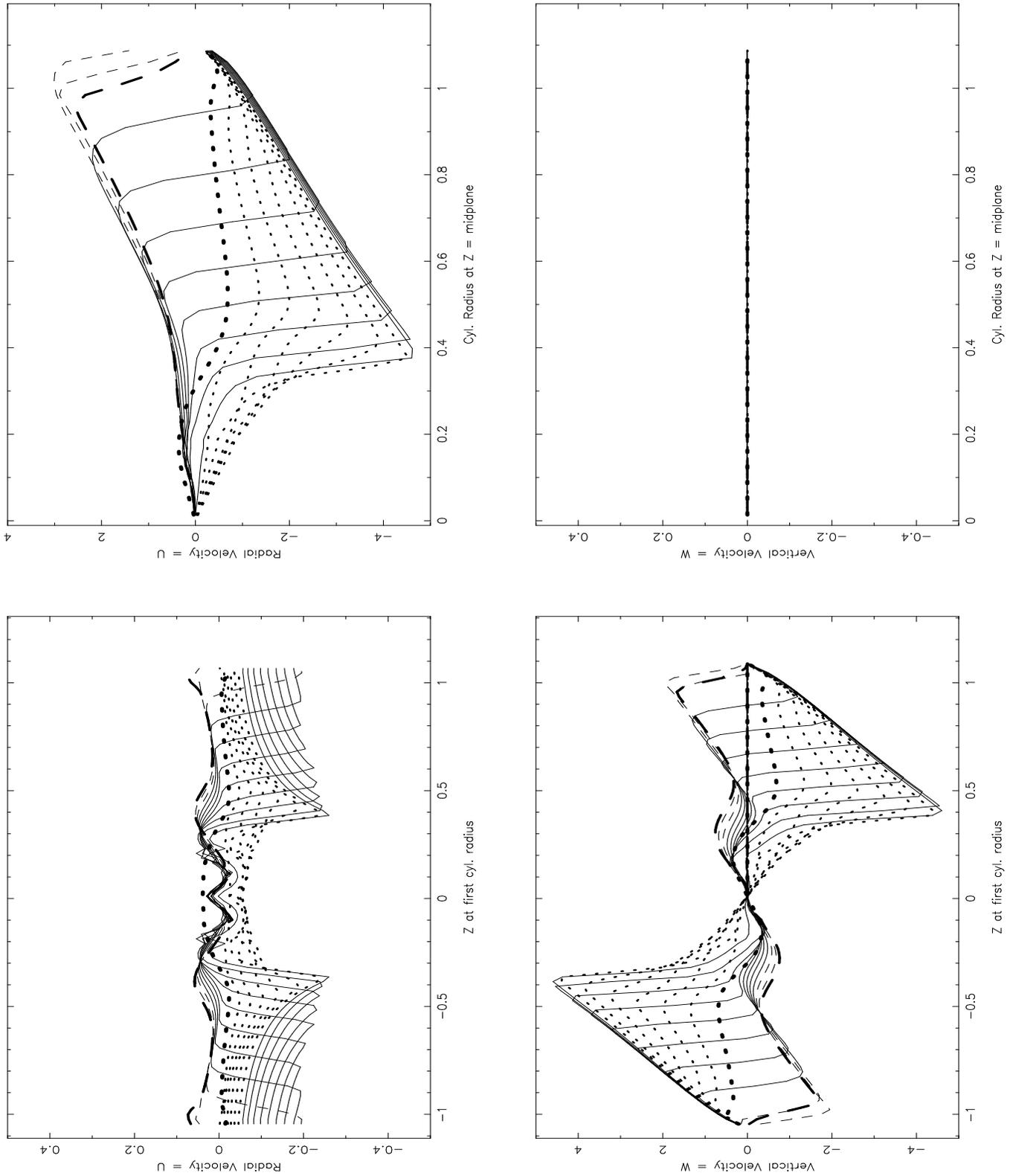}
  \caption[]{Time evolution of the radial velocity (top) and vertical
velocity  (bottom) along
two  rays of the cylindrical grid.  (cf. fig.~\ref{crossectdenslam})} 
  \label{crossectvels}
\end{figure*}

Figs.~\ref{crossectdenslam} and~\ref{crossectvels} show values of
the density, the inverse of the
temperature (lambda), and both radial and vertical velocity along two
rays through the cylindrical grid. 
One ray goes through a point $(\varpi=0,\phi=0,z=0)$ and extends
in the $\varpi$ direction.  We plot various quantities as a function
of $\varpi$ along this ray at time $t_{n}$.  The second ray goes through
a point $(\varpi=\varpi_{1},\phi=0,z=-1.1)$, extends in the $z$ direction
and we plot various quantities as a function of $z$ along this ray
at time $t_{n}$.  Thus in figs.~\ref{crossectdenslam} and~\ref{crossectvels}
each curve represents results
at one instant (specified by $t_n$) in the course of a run.  The curves are approximately
evenly spaced in machine time units.  The key for the lines styles
can be found in the right hand corner of the density plot,
(Fig.~\ref{crossectdenslam}).  In the following discussion we will
refer primarily to these two figures (~\ref{crossectdenslam} 
and~\ref{crossectvels}) and to the times
for which information is presented in these figures. For a view
of the gas evolution in the entire $\varpi - z$ plane, as opposed
to merely along these two rays, we will also refer to a time 
sequence of density
and lambda surface plots, as well as to plots of velocity vectors 
(Figs.~\ref{1surfixpot},~\ref{2surfixpot},~\ref{3surfixpot}).  In the surface
plots, values of density and lambda are given  on the
vertical axis as a function of $\varpi$ and $z$
in the horizontal plane.  In the velocity plots, each arrow represents
the vector sum of the radial and vertical velocities at a point on the
$\varpi - z$ plane.  The length of the arrow is proportional to the
strength of the velocity field.

From $t_{1}$ through $t_{7}$, the gas density
(Fig.~\ref{crossectdenslam}) rises at the centre
of the grid, although the grid centre is not where it reaches its
highest values initially (see also Fig.~\ref{1surfixpot}).  This is 
because the strong gravitational 
forces (Fig.~\ref{fixgforce}) keep the gas
from falling uniformly into the gravitational potential well. 
Instead the gas density piles highest along
a ring surrounding the grid's centre.  The peak of the ring moves inward from
$t_{1}$ through $t_{7}$, and the density jump between the tip
of the ring and the region outside of it is steepest and reaches a maximum at
$t_{8}$ in these figures (see first density plot in
Fig.~\ref{2surfixpot}). Time $t_{8}$ marks the onset of the
outgoing shock into the cold, low-density region outside the ring.

The lambda plots (Fig.~\ref{crossectdenslam}) show that between 
$t_{1}$ and $t_{7}$, the
region outside the ring becomes progressively colder.  This is a 
consequence of the regions nearer to the ring falling inward more
quickly then the regions further from the ring
(Fig.~\ref{crossectvels}), leading to an
expansion of the gas and a concurrent cooling.  By $t_{7}$ the leading
edge of the inwardly spreading
cold region has steepened into a shock. From $t_{8}$ through
$t_{15}$ the heated centre expands sending an outgoing shock which
heats the gas as it propagates across the grid.  We vividly see both the
formation of the shock and its outward movement
in the lambda surface
plots (Figs.~\ref{1surfixpot},~\ref{2surfixpot} and
~\ref{3surfixpot}).  In these surface plots, the cooling of the 
regions further from the ring is also visible.

The formation of the shock is also clearly seen in the
velocity profiles (Fig.~\ref{crossectvels}) 
from $t_{1}$ through $t_{7}$.  During this time interval, the
point of maximum velocity moves inward until at $t_{8}$ the ring
material starts moving outward and encounters the material still
infalling from the outer part of the grid at a shock front. 
From $t_{8}$ to $t_{15}$ we see the shock propagating outward.
The outwardly moving shock front is delineated clearly in the 
velocity vector plots (Figs.~\ref{1surfixpot},~\ref{2surfixpot} and
~\ref{3surfixpot}).

\begin{figure*}
\rotate[l]{\epsfxsize=15.cm \epsfysize=17.cm \epsfbox{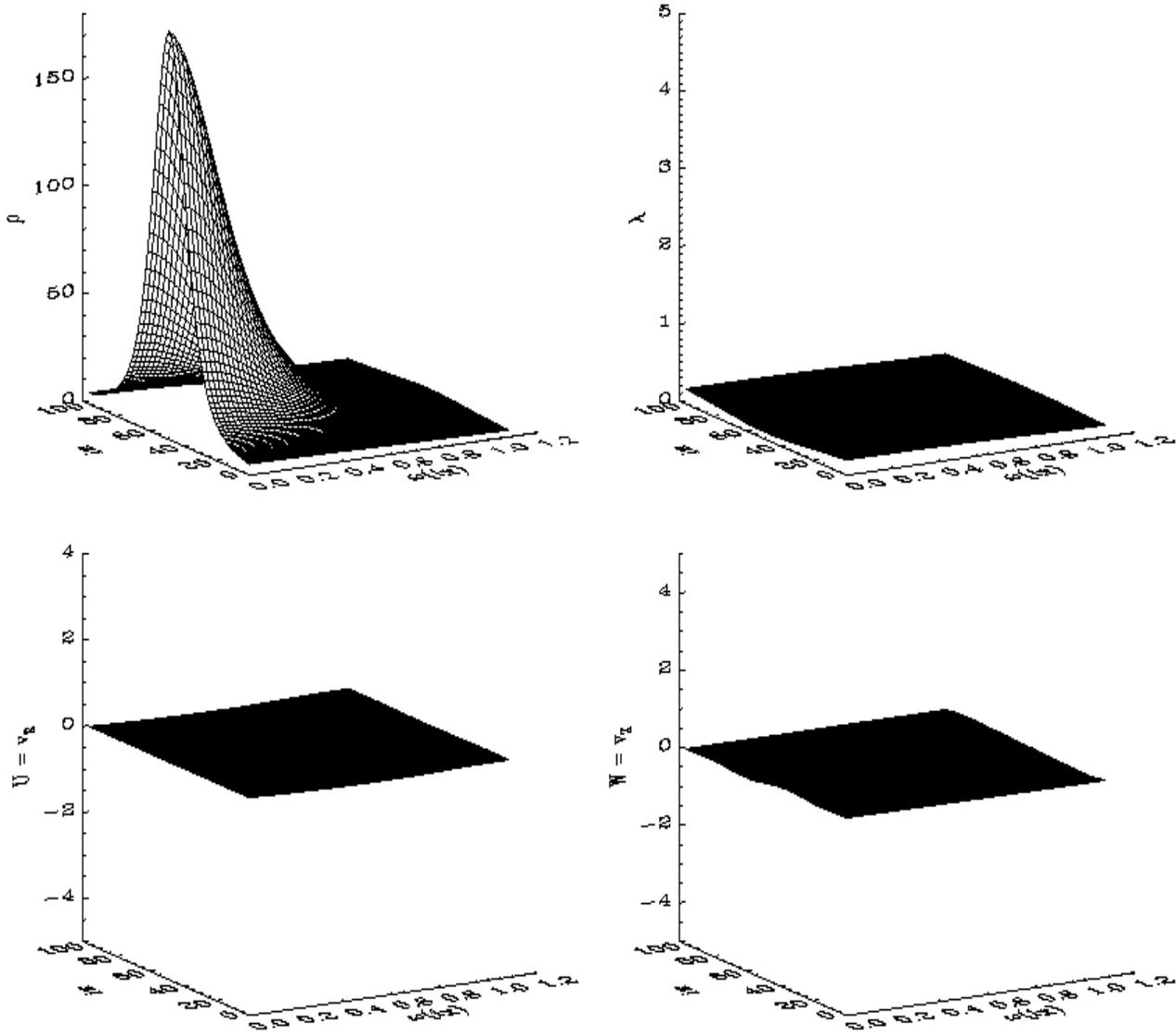}}
  \caption{Density (top left), lambda (top right), radial velocity
(bottom left) and vertical velocity (bottom right) after 100,000
  iterations which is equivalent to a total time in machine units
of 36.3758. This amounts to about 560 free-fall times.  Note that
lambda is nearly constant and the velocity in both $\varpi$ and $z$ is
nearly zero, indicating that the code has (nearly) reached the 
expected equilibrium state.} 
  \label{fixpotfinal}
\end{figure*}

After $100,000$ iterations (Fig.~\ref{fixpotfinal}), the gas is almost
in equilibrium: it has 
a smooth density profile and it is nearly isothermal and stationary.
Computing the value of the free-fall time ($t_{ff}=\sqrt{\frac{3
\pi}{32 G \rho}}$) from the initial conditions, $t_{ff} \approx .06$.
Therefore at $100,000$ iterations (time $= 36.3758$) the time is $560$
in units of the initial free-fall time.

\begin{figure*}
\epsfxsize=18.0cm \epsfysize=21.cm \epsfbox{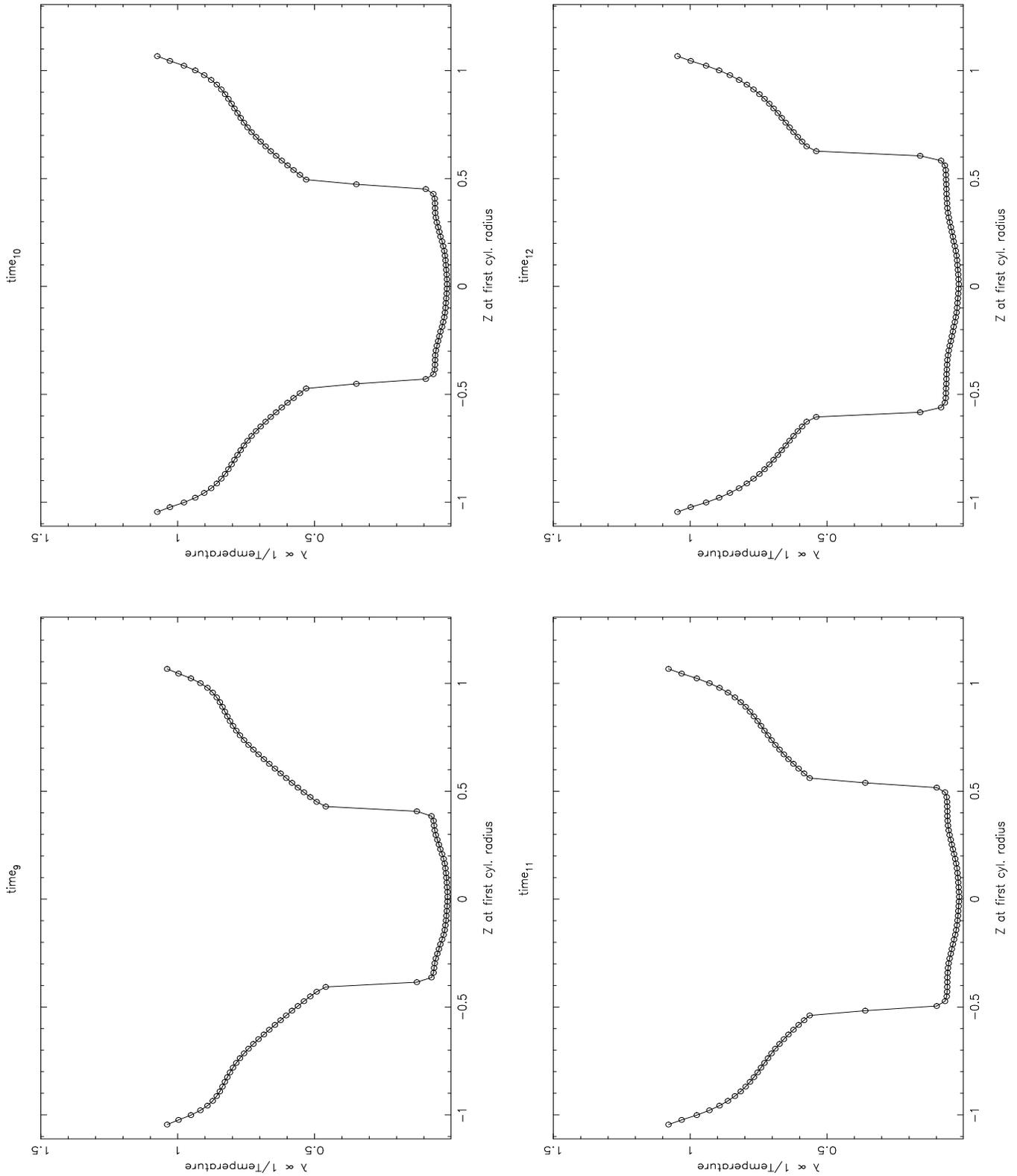}
  \caption{Lambda plotted once again at different $z$ along
  $\varpi_{1}$ for $t_{9} \rightarrow t_{12}$.  This time the
data for each of the four time instances is shown in a separate plot.
The circles are the values of $\lambda$ in each cell and the thin
  lines are curves through these values.  Note the sharpness of the
shocks and the absence of near-shock oscillations.  The shocks have not
been sharpened in any way, nor have any oscillations been suppressed;
all cells are shown in each plot.} 
  \label{shockres}
\end{figure*}

The most significant thing to notice in the results from this
computation are the sharp shock fronts. 
Even though the computation was performed
on a relatively coarse grid ($50$ cells in $\varpi$ and $100$ cells in $z$), 
only about
two cells are needed to resolve a shock and there is no evidence of
under
and overshooting.  This is shown clearly in Fig.~\ref{shockres},
where the inverse
of the temperature, $\lambda$, for $t_{9} \rightarrow t_{12}$,
is again plotted for the ray which cuts the cylindrical grid 
along the first cylindrical radius, $\varpi_{1}$ for fixed $\phi=0$.
However in Fig.~\ref{shockres}, we show the data for each of the four time
instances in a separate plot, so as to better assess the
shock resolution. The circles are the values
of $\lambda$ in each
cell and
the thin lines are curves through these values. It is easy to see
that only about two cells are needed to resolve a shock.

\section{Discussion}
\label{sec-discussion}
The most notable results of the preceeding test runs may be summarized
as follows:
\newline \noindent 1.)For a fixed, external gravitational field both
in Cartesian geometry and in axisymmetric cylindrical
geometry with a conservative form for the energy equation,
the BGK scheme is able to keep a configuration in 
hydrostatic equilibrium for many sound crossing times. 
Furthermore with the energy equation in conservative form the 
BGK scheme settles to the correct physical solution.

\noindent 2.)The BGK scheme is able to 
 achieve high resolution with rather coarse grids. 
For example the simulation of gas falling into a fixed external
spherically symmetric gravitational potential uses $50$ cells in
$\varpi$ and $100$ cells in $z$ and achieves shocks whose fronts span
only $1$ or $2$ cells.  We
never resort to regridding in the neighborhood of a discontinuity
to achieve high resolution.
Conventional hydrocodes frequently strive for high resolution 
through grid refinement techniques. This approach is both
expensive and ineffective beyond  a certain point.  
Uniform grid refinement shortens time steps: in $3$ dimensions, a decrease by 
a factor of $2$ in the grid spacing of each of the $3$ dimensions involves
a factor of $16$ increase in the number of cycles that the code has to
execute to reach the same total time.  Furthermore even if one is
willing to accept the expense of short timesteps, no matter
how fine the grid, a diffusive scheme
never achieves the accuracy of a high resolution scheme run on the
same sized grid. With that said, local adaptive grid refinement 
for the BGK scheme (without gravity) has been implemented by 
Kim \& Jameson (1998) and
it gives excellent results for a number of tests involving unsteady
supersonic flows.

\section{Conclusion}

We incorporated gravity into the BGK scheme for hydrodynamics
and presented results for the case of time-independent gravitational
potentials both on a one-dimensional Cartesian grid and
on an axisymmetric cylindrical grid.  
Our results show that the BGK scheme is generalizable to the case
of a gas moving in the presence of an external, time-independent
gravitational field.  When we derive an approximate local solution
to the BGK equation, linear in space and time, 
gravity's curvature of particle trajectories is unimportant.
Gravity's effect on the flow enters into the computational method
only through gravitational source terms. 

We conclude by
stating that we believe that no other hydrodynamical scheme
which is used in astrophysical computational fluid dynamics
invokes as few ``fixes'' and operates with as much generality
as the BGK scheme.  There are no discontinuity detection algorithms, 
with subsequent special treatment of special regions.
An entropy fix never has to be invoked when following the
evolution of rarefaction waves because the BGK scheme satisfies
the ``entropy condition''.  Truly multi-dimensional BGK schemes (i.e.
non-operator split) are realizable.

We attribute the BGK scheme's performance to its derivation from a model
to the collisional Boltzmann equation.  A tremendous advantage of a hydrocode
designed on the basis of a solution to a model of such a fundamental equation
is that the scheme gets physical backing for its incorporation 
of viscous and heat conductive effects.  Velocity moments of a
time-dependent local solution to the BGK equation give hydrodynamical
fluxes which carry both advective and dissipative terms - the BGK
scheme does not decouple them into separate operations.  
Furthermore
the algorithm for solving the BGK equation in the
BGK scheme avoids the BGK equation's stiffness which would
cause the time step of the scheme to be the collision time of the gas
instead of the much larger CFL time.  Without such an algorithm to
bypass the BGK equation's stiffness, the timestep would be
prohibitively small for performing hydrodynamical simulations
in real time. 

While it may seem complicated to solve for fluid
evolution by following a changing distribution function (including the
effects of collisions) in phase
space, we believe that it is not.
In spite of its looks, 
the BGK scheme is a computationally straightforward explicit scheme, 
and all parts of the algebra are in explicit closed form
for a perfect gas.  Even when a matrix is inverted for the connection
between the moments and parameters of a Maxwellian, the inverse is 
explicitly known.  Unless readers were told they might not guess
from inspection of the code that it originated in gas-dynamic
considerations.  The code itself is just straightforward algebra.
There are no iterative steps in the
BGK scheme.  There is no Riemann solver, either exact or approximate.
No numerical sub-steps (e.g. Runge-Kutta steps) are used.

Thus far we have not been concerned with the optimization of
the code. Code development has focused on demonstrating that
the code correctly simulates physical phenomena.  Kim and Jameson
(1998) compared the CPU time required just for the BGK scheme's 
flux computation to the flux computation in two other high resolution 
schemes which use
flux-splitting methods: characteristic splitting using Roe
averaging (Csplit) (Jameson 1996) and CUSP(Convective Upwind Split
Pressure) splitting by Jameson (Kim and Jameson 1995).  They find that
the BGK scheme is less than twice as slow
as the two other schemes.  
Note that in contrast to the flux computation in the other schemes, 
the BGK flux computation includes extra 
arithmetical operations for Navier-Stokes terms.
When the other schemes include Navier-Stokes terms, they
are not spared this computational expense - it just appears 
outside of the flux computation.

\begin{acknowledgements}
A.S. acknowledges partial support from a NASA
Space Grant administered through Cornell University.
\end{acknowledgements}

\clearpage

\clearpage
\clearpage
\clearpage
\clearpage
\clearpage
\clearpage
\clearpage
\clearpage

\clearpage

\clearpage

\clearpage

\clearpage
\clearpage
\clearpage
\begin{figure*}
\epsfxsize=18.0cm \epsfysize=21.cm \epsfbox{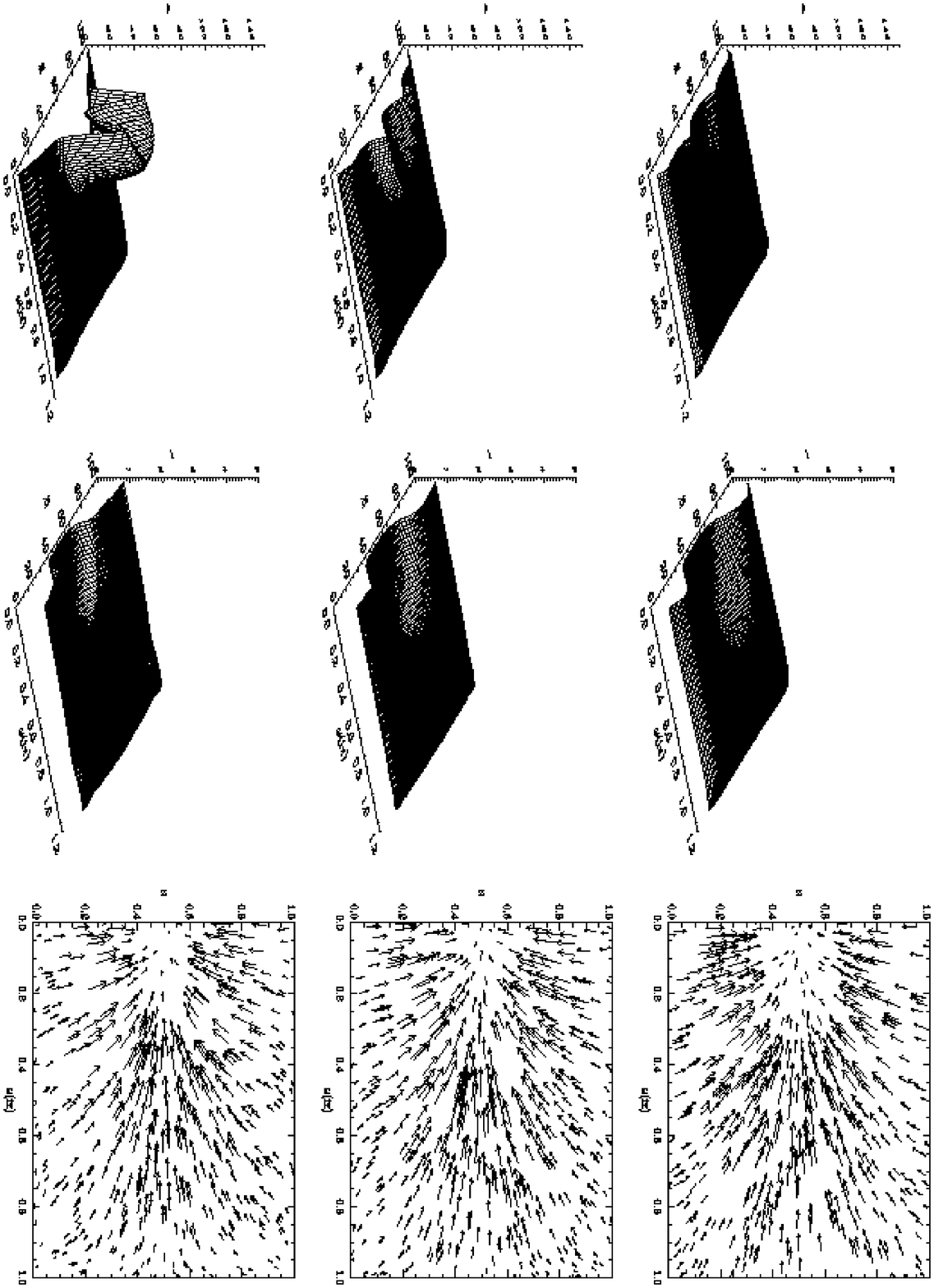}
  \caption[]{Time evolution of the density (first column), lambda
  (second column) and velocity (third column) for the following
  sequence of times (in units of the initial free-fall time): 
3.802D-03, 1.2671 , 2.723.} 
  \label{1surfixpot}
\end{figure*}
\clearpage
\begin{figure*}
\epsfxsize=18.0cm \epsfysize=21.cm \epsfbox{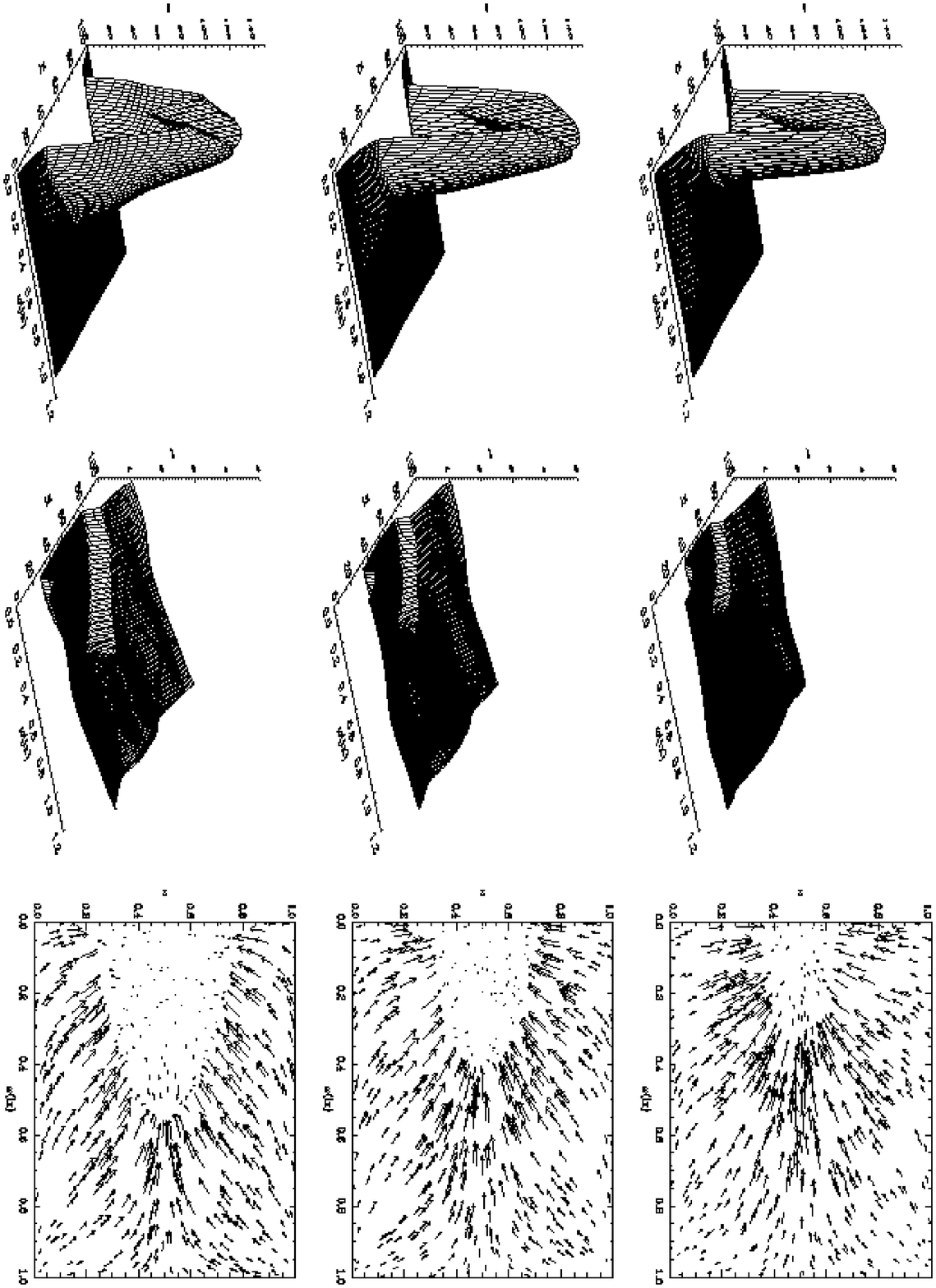}
  \caption[]{Continuation of time evolution of the density (first column), lambda
  (second column) and velocity (third column) for the following
  sequence of times (in units of the initial free-fall time): 
3.607, 4.8, 5.805.} 
  \label{2surfixpot}
\end{figure*}

\clearpage

\begin{figure*}
\epsfxsize=18.0cm \epsfysize=21.cm \epsfbox{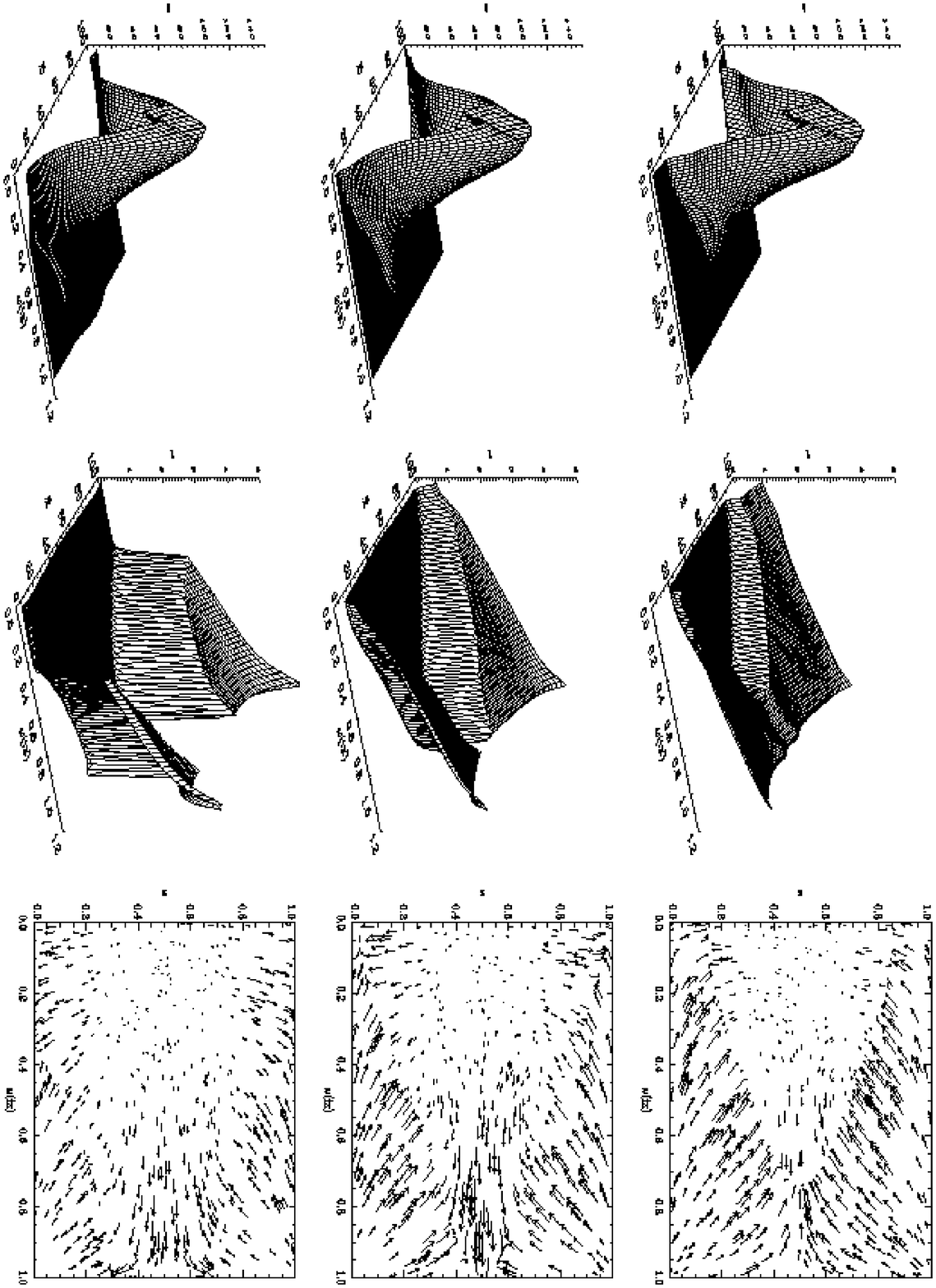}
  \caption[]{Continuation of time evolution of the density (first column), lambda
  (second column) and velocity (third column) for the following
  sequence of times (in units of the initial free-fall time): 
6.898, 8.067, 8.88.} 
  \label{3surfixpot}
\end{figure*}

\clearpage

\clearpage

\end{document}